\documentclass[a4paper,fleqn]{cas-dc}
\usepackage[numbers]{natbib}
\usepackage{float}
\usepackage{graphicx}
\usepackage{blindtext}
\usepackage[export]{adjustbox}
\usepackage{multicol}
\usepackage[utf8]{inputenc}
\usepackage{amssymb,amsfonts}
\everymath{\displaystyle}
\usepackage[many]{tcolorbox}
\usepackage[figuresright]{rotating}
\usepackage{array}
\usepackage{kantlipsum}
\usepackage{gensymb}
\usepackage{xpatch}
\usepackage{lipsum}
\usepackage{multirow} 
\usepackage{lineno}
\usepackage{amsmath}
\usepackage{amsbsy,enumerate}
\usepackage{ccaption}
\usepackage{chemformula}
\usepackage{balance}
\usepackage{geometry}
\usepackage{ragged2e}
\usepackage{setspace}
\usepackage{subcaption}
\usepackage{etoolbox}
\usepackage{siunitx}
\usepackage{mathtools}
\usepackage{xcolor}
\usepackage{fullpage,anysize,esint,relsize, commath}
\usepackage{fancyhdr} 
\usepackage{hyperref}
\usepackage{enumitem}
\usepackage{pifont}
\usepackage{listings}
\usepackage{cprotect}
\usepackage{multiaudience}
\usepackage{abbrevs}
\usepackage{isomath}
\usepackage{graphics}

\DeclareMathAlphabet{\mathpzc}{OT1}{pzc}{m}{it}
\DeclareMathAlphabet{\mathbbm}{U}{bbm}{m}{n}

\def\tsc#1{\csdef{#1}{\textsc{\lowercase{#1}}\xspace}}
\tsc{WGM}
\tsc{QE}
\tsc{EP}
\tsc{PMS}
\tsc{BEC}
\tsc{DE}
\begin{document}
\let\WriteBookmarks\relax
\def\floatpagepagefraction{1}
\def\textpagefraction{.001}
\title [mode = title]{Deep Multi-Task Learning for Malware Image Classification}
\author[1]{Ahmed Bensaoud}
\cormark[1]
\ead{abensaou@uccs.edu}
\author[1]{Jugal Kalita}
\ead{jkalita@uccs.edu}
\address[1]{Deptarment of Computer Science, University of Colorado Colorado Springs}
\cortext[cor1]{Corresponding author:}
 
\begin{abstract}
Malicious software is a pernicious global problem. A novel multi-task learning framework is proposed in this paper for malware image classification for accurate and fast malware detection. We generate bitmap (BMP) and (PNG) images from malware features, which we feed to a deep learning classifier. Our state-of-the-art multi-task learning approach has been tested on a new dataset, for which we have collected approximately 100,000 benign and malicious PE, APK, Mach-o, and ELF examples. Experiments with seven tasks tested with 4 activation functions, ReLU, LeakyReLU, PReLU, and ELU separately demonstrate that PReLU gives the highest accuracy of more than 99.87\% on all tasks. Our model can effectively detect a variety of obfuscation methods like packing, encryption, and instruction overlapping, strengthing the beneficial claims of our model, in addition to achieving the state-of-art methods in terms of accuracy.
\end{abstract}

\maketitle
\section{Introduction}
The number of attacks on computers and computer networks is rising all over the world. Malicious software (or malware) is everywhere, with people trying to steal information from commercial or non-profit organizations and/or governments, and benefit politically, financially or otherwise. For example, a report from Cybersecurity \& Infrastructure Security Agency (CISA) and Federal Bureau of Investigation (FBI) in May 2020 warned US organizations performing research on COVID-19 vaccines that foreign governments were attempting to hack into their system\footnote{\url{https://us-cert.cisa.gov/china}}. McAfee Labs reported that 419 threats per minute were observed in Q1 2021, an increase of almost 6.33\% over the previous quarter \cite{report}. Malware is a prominent threat to smaller systems as well. Malware has also shown up in smartphones using Android and iOS systems due to the downloading of thousands of applications (apps) from the Internet. Every smartphone vendor has an application market for its OS, including Google Play, Blackberry App World, Windows Phone Marketplace (Microsoft Azure), and Apple Store. Apple apps for iOS devices such as iPhone, iPad, and iPod Touch can be installed only from the proprietary Apple App Store. If iOS users want to install apps that have not been approved by Apple, they need to remove the manufacturer's restrictions by jailbreaking. A jailbroken device allows users to gain full access to the root of the OS and gives users additional control. One of the top risks associated with jailbroken devices is higher susceptibility to malware. Android allows users to install from outside Google’s app store, without jailbreaking. Android OS, which is open-source, is usually the first target of anyone who wants to develop malicious apps (see Fig. 1).
\begin{figure} [ht]
\centering
	\includegraphics[width=8cm, height=5cm, frame]{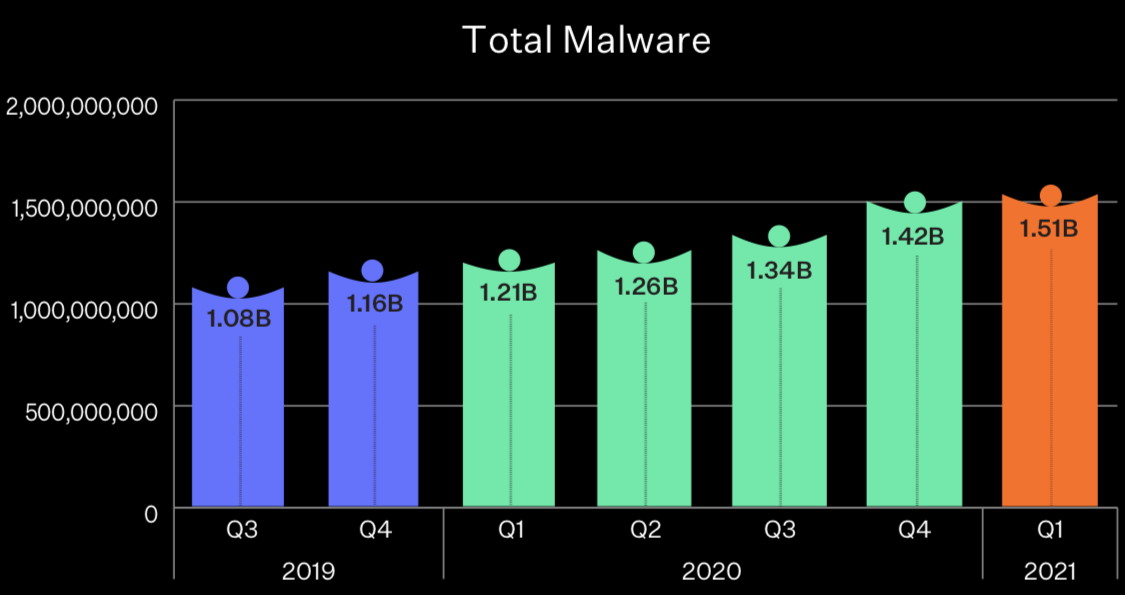}
	\caption{Total number of malware is increasing from quarter to quarter}
\end{figure}

McAfee Labs Threats Report: June 2021

It is often in the third-party app repositories where malware authors upload their software with a goal to enable hackers to take control of a device by stealing passwords, or providing access to contacts. Thus, the development of intelligent techniques for malware detection is an urgent need. Unfortunately, malware classification is still challenging even though current state-of-the-art classifiers have achieved excellent results in general, especially in computer vision. To support efficient and effective malware classification, we propose a multi-task learning model in this paper.\par 
Research on malware detection usually starts by extracting features from certain sections of malware files. In our research, we extract features from structural information in all files, in addition to performing dynamic analysis over the runtime behavior of the program files. We also extract features from the unpacked executables to detect obfuscation. To detect and classify malware using deep learning, we use all file sections instead of a specific section of the malware file like some previous efforts.\par 
Multi-task learning (MTL) has been used successfully in areas such as computer vision, natural language processing and speech processing. For example, MTL has been extended to deep learning to improve the performance in computer vision. We employ multi-task learning for binary and multi-class malware image classification using seven submodels.\par
The availability of datasets to test and evaluate proposed malware detection models has been a bottleneck. That is why we create a large modern dataset as well. We make this dataset available to researchers.\\
Our contributions in this work are:\par
\begin{itemize}
\item We propose and implement a novel multi-task learning architecture for deep learning for malware classification.
\item We create a benchmark color image dataset representing malware from portable executable, Android, ELF, Mac OS, and iOS files, and make it publicly available for the benefit of the research community\footnote{\url{https://github.com/abensaou-uccs/Colorado-MalColorImg}}.
\item We conduct extensive experiments with various multi-task learning architectures for deep learning. Experimental results show that the proposed framework obtains excellent classification performance, achieving average accuracy of 99.97\%. Our model is the state-of-the-art in malware classification. 
\end{itemize}

The following is how we organize the paper's overall structure. Section 2 discusses malware detection methods and multi-task learning. Section 3 details our methodology. Section 4  describes our method for generating image-like files from malware files to facilitate classification. Section 5 shows how we generate additional Mac OS malware samples using CycleGAN because the original numbers are small. Section 6 outlines our proposal of multi-task learning. Section 7 describes all datasets. In Section 8, we perform classification and evaluate the model. Section 9 presents experimental results. Finally, we provide conclusions and present the limitations of the study in Section 10.
\section{Related work}
This section discusses the related work regarding  malware detection approaches, malware visualization and classification based on deep learning, and multi-task learning.

\subsection{Malware detection}
 To defend computer systems from malware, we need to detect malware before it affects the computer systems. Three traditional approaches have been used to detect malware:
Signature-based detection, Heuristic detection, and Behavior-based  detection. These methods have a number of advantages and disadvantages. 1) Signature-based detection efficiently identifies known malware using pattern matching, but is unable to identify unknown malware since malware can change its characteristics, producing a new signature that signature-based detection cannot identify. 2) Heuristic detection can identify known and unknown malware, but this method can lead to high error rates for false-positives and false-negatives. 3) Behavior-based malware detection approaches observe the behavior and purpose of a suspicious file. This approach needs resources and time to execute and monitor the behavior of the suspicious file.\\ 
Machine learning models have also been applied in malware detection. Supervised machine learning algorithms such as naive Bayes (NB), C4.5 decision tree variant (J48), random forests (RF), support vector machines (SVM), sequential minimal optimization (SMO), k-nearest neighbors (KNN), multilayer perceptron (MLP), and simple logistic regression (SLR) have been used to detect malware. \citet{ucci2019survey} surveyed malware analysis using machine learning techniques and discussed various features that have been used by researchers to improve malware detection systems. Recently, researchers have also attempted to detect malware using deep learning. They have indicated that basic deep learning models perform well in malware analysis. In addition, researchers have shown that it is possible to build a combination of two or more models to strengthen the outcomes.\\
\citet{ alzaylaee2020dl} proposed DL-Droid, a deep learning system to detect malicious Android applications through dynamic analysis. They evaluated the model using 31,125 Android applications, 420 static and dynamic features, and compared performance with existing DL-based frameworks. Dynamic features on DL-Droid achieved up to a 97.8\% detection rate and 99.6\% with dynamic and static features. 
\citet{darabian2020detecting} used static and dynamic analysis for system calls of Portable Executable (PE) samples of cryptomining applications. They collected system call data and then fed into the deep learning models---LSTM, Attention-based LSTM, and Convolutional Neural Networks (CNNs). Their models achieved 95\% accuracy rate in static analysis on opcode and accuracy rate of 99\% in dynamic analysis on system calls.
Deep learning models for static and dynamic analysis of malware have been explored, producing promising results to detect obfuscated malware \cite{souri2018formal}. 

Malware authors develop obfuscation techniques such as packaging, shuffling, encryption, and tokenization to make it harder to detect, thus, elude  anti-malware engines \cite{SIHAG2021100365}. The current  obfuscation detection techniques for Android  applications perform poorly. A recent survey on Android malware detection using the latest deep learning algorithms has investigated the challenges and analyzed the results of obfuscation detection systems\cite{qiu2020survey}. In addition, detection of Linux malware is also still in its infancy, but such malware is already using IoT devices of different negative behaviors and tricks. Unfortunately, analyzing ELF files is quite difficult since Linux runs in devices of all kinds from a really small tier masters to really large servers \cite{NGO2020280}.
\subsection{Malware Image}

Malware executable can be represented as a matrix of hexadecimal or binary strings and converted to a form which can be thought of as an image. To create a new malware, malware authors usually add to or change the code in old malware. Thus, when viewed as an image, one can easily visualize small adds or changes in various sections of the file structure. \citet{nataraj2011malware} first proposed a technique to convert malware into images, transforming the raw bytecode PE files to greyscale image data where a pixel is represented by a byte.
\citet{vasan2020image} performed image-based malware classification using an ensemble of CNN architectures to detect packed and unpacked malware files. The approach used two pre-trained models, VGG16 and ResNet-50, which were fine-tuned for classification of malware images. The approach achieved more than 99\% accuracy for unpacked malware and over 98\% accuracy for packed malware.

\citet{su2018lightweight} converted malware binary in IoT environments to image and used a light-weight convolutional neural network to classify malware families. Their model achieved 94.0\% accuracy for goodware and DDoS malware, and 81.8\% for goodware and two powerful malware families.

\citet{ni2018malware} proposed the MCSC (Malware Classification using SimHash and CNN) model which hashed decompiled malware code and converted it to grayscale images, and then trained CNNs for classification. Their model achieved an average accuracy of 98.86\% on a malware dataset of 10,805 samples.

\citet{bensaoud2020classifying} used six deep learning models for malware classification. Three of these are past winners of the ISVLVRC contest: VGG16, Inception V3, and ResNet50, and the other three models are CNN-SVM, GRU-SVM, and MLP-SVM, which enhance neural models with support vector machines (SVM). They trained all models on the Malimg dataset \cite{nataraj2011malware}, and the results indicate that the Inception-V3 model achieved a high test accuracy of 99.24\% among all compared work.

\citet{naeem2020malware} converted APK files to color images and then fed them to a DCNN model. The model achieved 97.81\% accuracy on the Leopard Mobile malware dataset\footnote{\url{ https://sites.google.com/site/nckuikm/home}} and 98.47\% accuracy on a Windows dataset\footnote{\url{https://vision.ece.ucsb.edu/research/signal-processing-malware-analysis}}.

\citet{kalash2018malware} proposed a CNN-based architecture to classify malware samples. They performed experiments on two datasets, Malimg and Microsoft malware. The method achieved 98.52\% on 9339 Malimg malware samples from 25 malware families and 99.97\% on 21,741 Microsoft malware samples\footnote{\url{https://www.kaggle.com/c/malware-classification}}.

\citet{mercaldo2020deep} proposed a platform for a static approach for the detection of malicious samples using supervised deep learning. They gathered a set of features from gray-scale images to build several classifiers to identify the belonging malware family and the variant inside the family. The family detection model obtained an accuracy of 93.50\% and variant detection obtained an average accuracy of 95.80\% 

\subsection{Multi-Task Learning}
Often learning how to perform several tasks simultaneously helps perform one or more of the tasks better than learning to perform tasks individually. We use multi-task learning in our work since it can be used for binary and multi-class malware classification simultaneously. It can also improve the performance on several related datasets.\\

In multi-task learning, multiple related tasks are learned jointly, and useful information is shared among related tasks. Each task benefits from other tasks producing better results for one or more trained tasks. Several related tasks are learned jointly from a shared dataset. There are two approaches, 1) Hard parameter sharing, where hidden layers are shared by all tasks, with different output layers as shown in Fig. 2. 2) Soft parameter sharing, where different tasks have their own networks, but parameters are made similar by regularization, and the output layers are different, as shown in Fig. 3. One important advantage of hard parameter sharing is that it reduces the number of parameters in the model since the same feature space is used by all closely related tasks. In addition, it performs as a regularizer that reduces the risk of overfitting and makes the model architecture compact for efficient training. Moreover, hard sharing parameters use a single shared representation and then connect to multiple tasks, with each task represented as a submodule. \citet{meyerson2018pseudo} showed that combining gradients from each task improves learning. \citet{zhang2020deep} proposed a deep learning-based multi-task learning approach to predict network-wide traffic speed, using a set of hard parameters and Bayesian optimization.

\begin{figure} [ht]
	\includegraphics[width=\linewidth, frame]{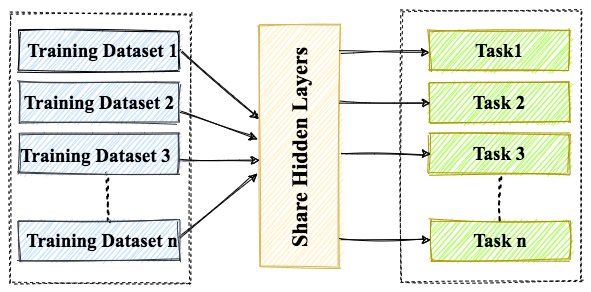}
	\caption{Hard parameter sharing for multi-task learning (MTL)}
\end{figure}

\begin{figure} [ht]
	\includegraphics[width=\linewidth, frame]{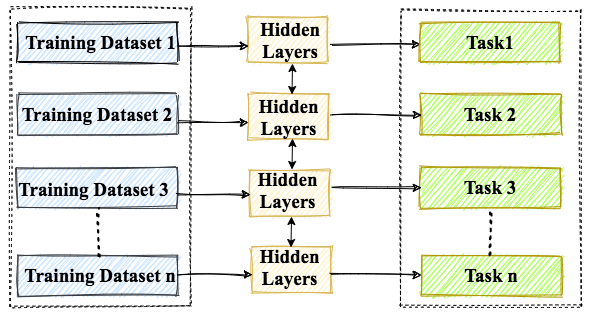}
	\caption{ Soft parameter sharing for multi-task learning (MTL)}
\end{figure}

Many MTL methods have been proposed to solve classification problems in computer vision. \citet{gkioxari2014r} applied a convolutional neural networks for the tasks of pose prediction and action classification of people in unconstrained images. \citet{zhao2020attribute} proposed an attribute hierarchy based multi-task learning (AHMTL) approach for fine-grained image classification on CUB-200-2011 \cite{wah2011caltech} and Cars-196 datasets \cite{krause20133d}. Their approach achieved the best classification performance at the time. 
\citet{bell2020groknet} built an image product recognition system called GrokNet, using a multi-task learning approach, matching the accuracy of a previous state-of-the-art Facebook product recognition system.
\citet{yu2020bdd100k} constructed BDD100K, a large-scale driving video dataset with 100K videos and 10 tasks for image recognition algorithms on autonomous driving. They created a benchmark for heterogeneous multitask learning and studied how to solve the tasks together. The results showed interesting findings about allocating the annotation budgets in multitask learning. 
\citet{chen2020spatial} proposed a multi-task learning based salient region detection method by fusing spatial and temporal features. The model learned a two-stream Bayesian model by integrating spatial and temporal features in a unified multi-task learning framework, outperforming previous methods.
\citet{wang2020towards} used a semi-supervised learning technique to address the missing visual field measurement label problem in the training set, and built a multi-task learning network to explore the relationship between the functional and structural changes in glaucoma and classify optical coherence tomography (OCT) images into glaucoma and normal. They achieved good results for the automated diagnosis system.
\citet{dorado2020multi} proposed and evaluated a multi-task deep neural network architecture for predicting Wind Power Ramps Events (WPREs) in three different classes. They modified the Adam optimization algorithm for imbalanced data for the misclassified class. Their model achieved very good performance for all the classes.

\section{Methodology}
We apply a multi-task learning model to learn from several malware datasets. The datasets are built with malware for PE Windows, APK Android, ELF Linux, and Mach-O for MacOS X. We extract features from sections in each malware file and convert them to RGB images and feed them to our model.

\subsection{PE Malware}
The Portable Executable (PE) file is the structure of all executable files (EXE) and Dynamic Link Libraries (DLL) that can be loaded and executed on any version of Microsoft Windows. The structure of a PE file includes DOS Header, PE Header, Optional Header, Sections Table, and sections that contain Code, Import, and Data. A PE file relies on several DLL files for execution, and each DLL is related to other DLLs to implement a certain task. 
The actual executable has different sections such as .text, .data, .idata, .edata, .rsrc, .reloc, .bss, and .debug, as shown in Fig. 4.
\begin{figure} [ht]
	\includegraphics[width=7.5cm, height=4cm, frame]{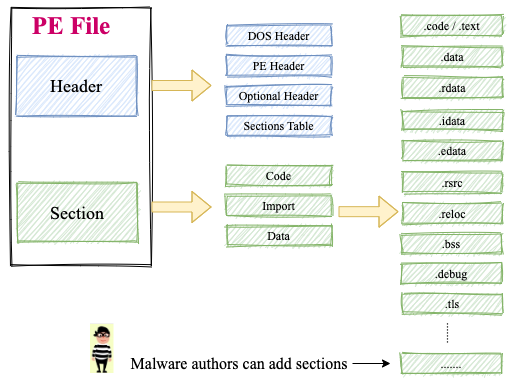}
	\caption{PE file structure}
\end{figure}

\subsection{Executable and Linkable Format (ELF)}
An executable file on Linux is structured using the ELF syntax. There are two different file formats, dynamic libraries (*.so) and  object  files  (*.o) \cite{drepper2006write}. Low focus on Linux threat detection by the antivirus industry motivates many malware writers to attack this operating system. Researchers have shown that ELF malware is quite complicated, demonstrating that malware authors put in a lot of effort in creating Linux malware. Examples of Linux malware are Rootkits, Exploits, and Backdoors, that DDoS attack malware usually use. The latest and most dangerous ELF malware is a combination of various individual types. We extract all ELF features from the samples: ELF headers, program header tables, code, data, section names, and section header tables. We use Pwntools, which is a convenient Python library to extract the hexadecimal form of malware ELF files. 

\subsection{MacOS X and iSO Malware}
The core of any operating system is known as the kernel. In MacOS and iOS, the name of the kernel is called Mach kernel, and its executable format is named  Mach-O file. There are several types of Mach-O files, such as .o, .dylib and .bundle extensions. We download the iOS App Store Package .ipa file from iTunes and remove digital rights management (DRM) protection to get decrypted executable. In addition, we collect malware samples from VirusTotal\footnote{\url{https://www.virustotal.com}} and Contagio\footnote{\url{http://contagiodump.blogspot.com}}. It is hard to find open source malware repositories from MacOS. This is the reason why we use the CycleGan technique to generate more Mach-O malware. For benign Mach-O files, we collect open source programs. The dataset consists of 5000 malware samples and 2000 benign samples. 

The Mach-O malware features are extracted from all section headers, load commands, and segments by parsing the file structures using a Python script, and are converted to hexadecimal as shown in Fig. 5. 

\begin{figure} [ht]
	\includegraphics[width=\linewidth,frame]{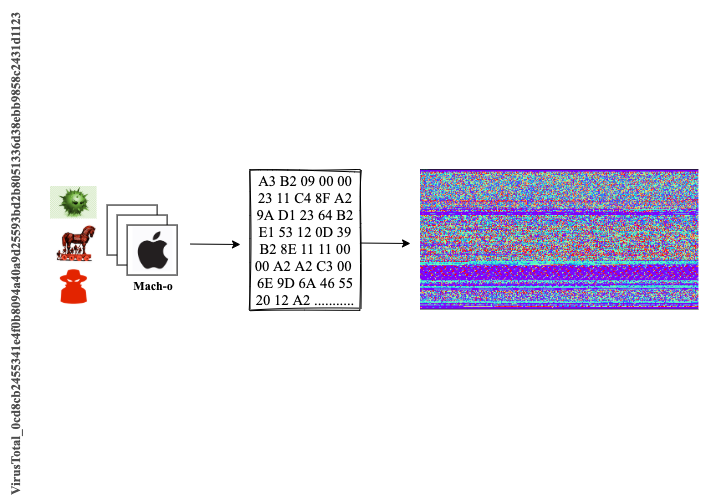}
	\caption{Converting MacOS malware Mach-o file to image}
\end{figure}

\subsection{Android Malware}
Android Package (APK) is a compressed (ZIP) bundle of files used by Android OS for mobile apps. We extract the contents of APK files: AndroidManifest.xml, classes.dex (interpreted by Dalvik VM), META-INF and resource files. 
The AndroidManifest.xml file describes information in the APK file like the package name, app components, the manifest permissions needed to access protected resources and hardware components. We parse both the Android manifest file and classes.dex to extract the features. We convert all these files to hexadecimal, merge them, and generate the RGB images using Android library in Python as shown in Fig. 6.
\begin{figure} [ht]
	\includegraphics[width=\linewidth,frame]{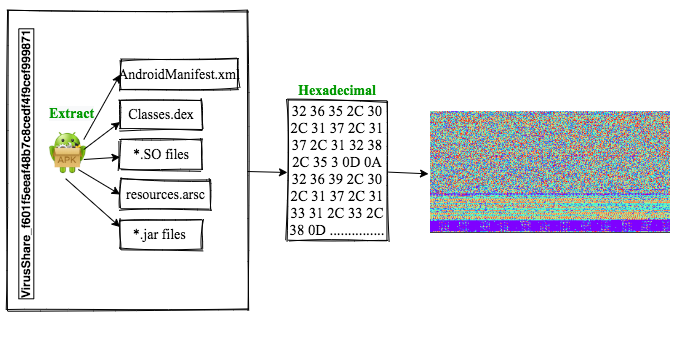}
	\caption{Converting Android malware APK file to image}
\end{figure}
\subsection{Unpacking Malware}
We consider malware packing since malware authors obfuscate, compress and encrypt malicious code via executable packing, making it hard to detect malware \cite{choi2011detecting}. We unpack malware using the state-of-the-art packing detection and packing classification tool PEiD \cite{PEiD}. The goal of unpacking malware is to merge the static and dynamic features in one image. 


\subsection{Assembling Code to Image}
We convert the malware samples to assembly code by using a popular disassembler, IDA Pro\footnote{\url{https://www.hex-rays.com/products/ida/}}. IDA Pro decodes binary machine code into readable assembly language code as shown in Fig. 7. 
\begin{figure} [ht]
	\includegraphics[width=\linewidth, frame]{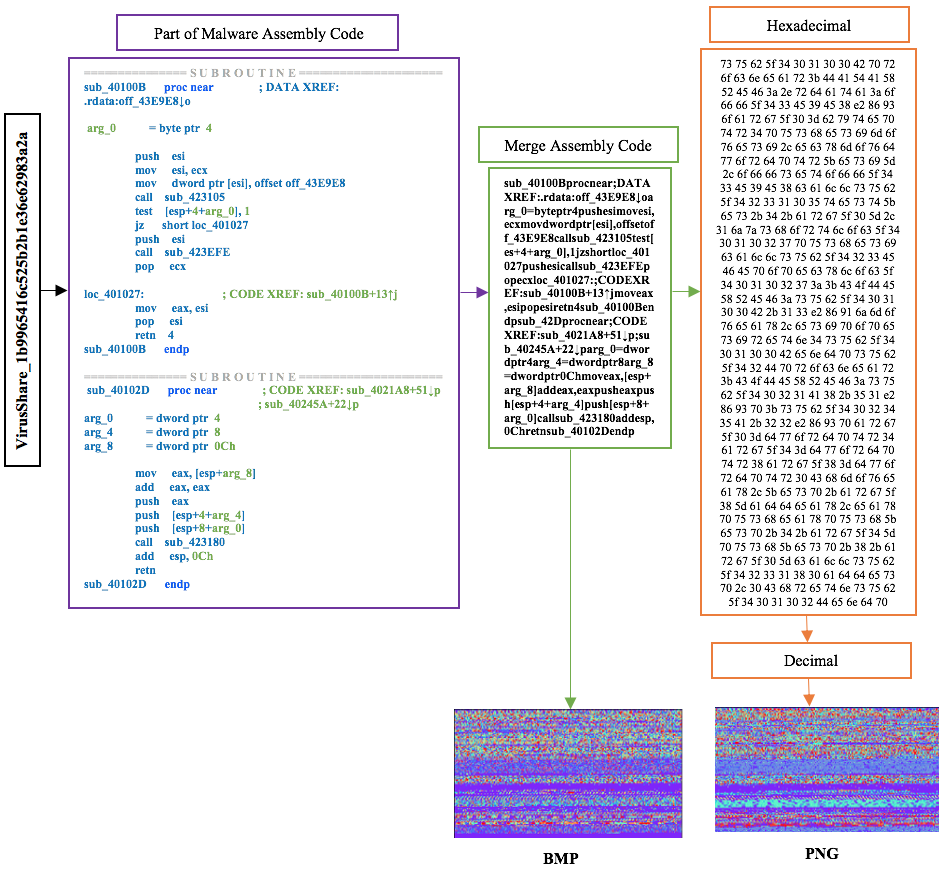}
	\caption{Converting sequence of assembly instructions of the malware to Image}
\end{figure}
\section{Generating Images}
An image is a set of numbers that record the intensity of red, green, and blue at every pixel location in an image in a grid format of pixels. Each pixel in the image can be represented by a vector in three primary color spaces, red (R), green (G), and blue (B). Let

\begin{equation*}
 \hspace{1cm} P_{i}= 
\begin{bmatrix}
R_{i} \\
G_{i} \\
B_{i}
\end{bmatrix}
\end{equation*}
where $P_i$ is the $i$th pixel of the image, $1\leq i\leq S$, where the size of the image is $S$. The mean value ($\mu$) and the standard deviation ($\sigma$) represent global features of the image and are calculated as follows:\\

$\mu=\frac{1}{S}\sum_{n=1}^{S} P_i,$\\

$\sigma=\sqrt{\frac{1}{1-S} {\sum_{n=1}^{S} {(P_i-\mu)^2.}}}$\\

Fig. 8 shows how an RGB pixel is stored in the image format. For instance, a pixel can be stored in up to 48 bits, with sixteen bits per channel.
\begin{figure} [ht]
	\includegraphics[width=\linewidth, frame]{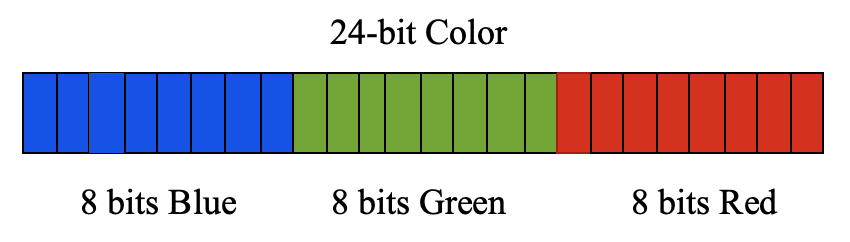}
	\caption{PNG 24 bits image in RGB color}
\end{figure}

\subsection{Malware Image Generation}
Files that look like images can be generated from a malware file in either raster or vector form. Raster images store the data as a grid of pixels. The most common raster image formats include Windows Bitmap (BMP), Joint Photographic Experts Group (JPEG), and Portable Network Graphics (PNG) images formats. On the other hand, vector format images are made up of objects based on geometric features such as circles, lines, polygons, rectangles, and curves. The most common vector images are Scalable Vector Graphics (SVG) and Vector Markup Language (VML). In this paper, we use BMP and PNG raster image formats since there is no need to scale up the images that SVG and VML vector formats allow.

In an RGB image, each channel's values go from 0 to 255, assuming each color channel is represented with 8 bits or a byte. The maximum number 8 bits can represent is 255 and the lowest is 0. Therefore, there are 256 possible intensities per color channel. For example, to represent a pixel of the color Sunglow, the values of RGB are 250 for a lot of red, 200 for a lot of green, and 62 for a little blue. The computer stores them as Red: 11111010, Green: 11001000, and Blue: 11111000, using 24 binary digits to represent such a pixel. We can represent the same color Sunglow using only six hexadecimal digits as FA C8 3E, which is a lot shorter than binary. A wide range of application software graphics packages express colors using hexadecimal codes. The image file can be stored in uncompressed (lossless) or compressed (lossy) formats. 

In addition, rows can be padded if necessary. For example, if we have an image size $17 \times 17$, each row needs $17$ times $3= 51$ bytes. We can test if the rows need padding or not by finding the modulo of row by $4$ $(51$ MOD $4= 3)$. The result of the modulo gives us how many padded rows are needed.
\subsection{Bitmap (BMP) and Portable Network Graphics (PNG)}

The bitmap (map or array of bits) file format is used to store  two-dimensional color images \cite{smart2005cross}. BMP files are uncompressed. They are also larger than other format images such as JPEG, GIF, and PNG. The bitmap file format has sections which contain file header, information header, data color table, and pixel values. A bitmap contains precise information about each and every pixel. BMP is the best type of input for malware classification using image files since the format does not use compression. A key reason for using large file sizes is to keep all malware features. On the other hand, the PNG file format uses lossless compression to store raster images in smaller space. Similar to BMP, PNG also keeps all malware features.

 When we generated a BMP or PNG image from malware file as shown in Fig. 9, we had to consider padding since there is a gap between row width and length of the image. For example, if the image is $51\times51$ with depth 24-bit, each row needs $51\times3=153$ bytes. We also compressed the BMP image using Run Length Encoding, which is a lossless compression approach. We use this approach since it reduces the size of the malware BMP image with no loss of information (see Figs. 10 and 11).

\begin{figure} [ht]
		\includegraphics[width=\linewidth, frame]{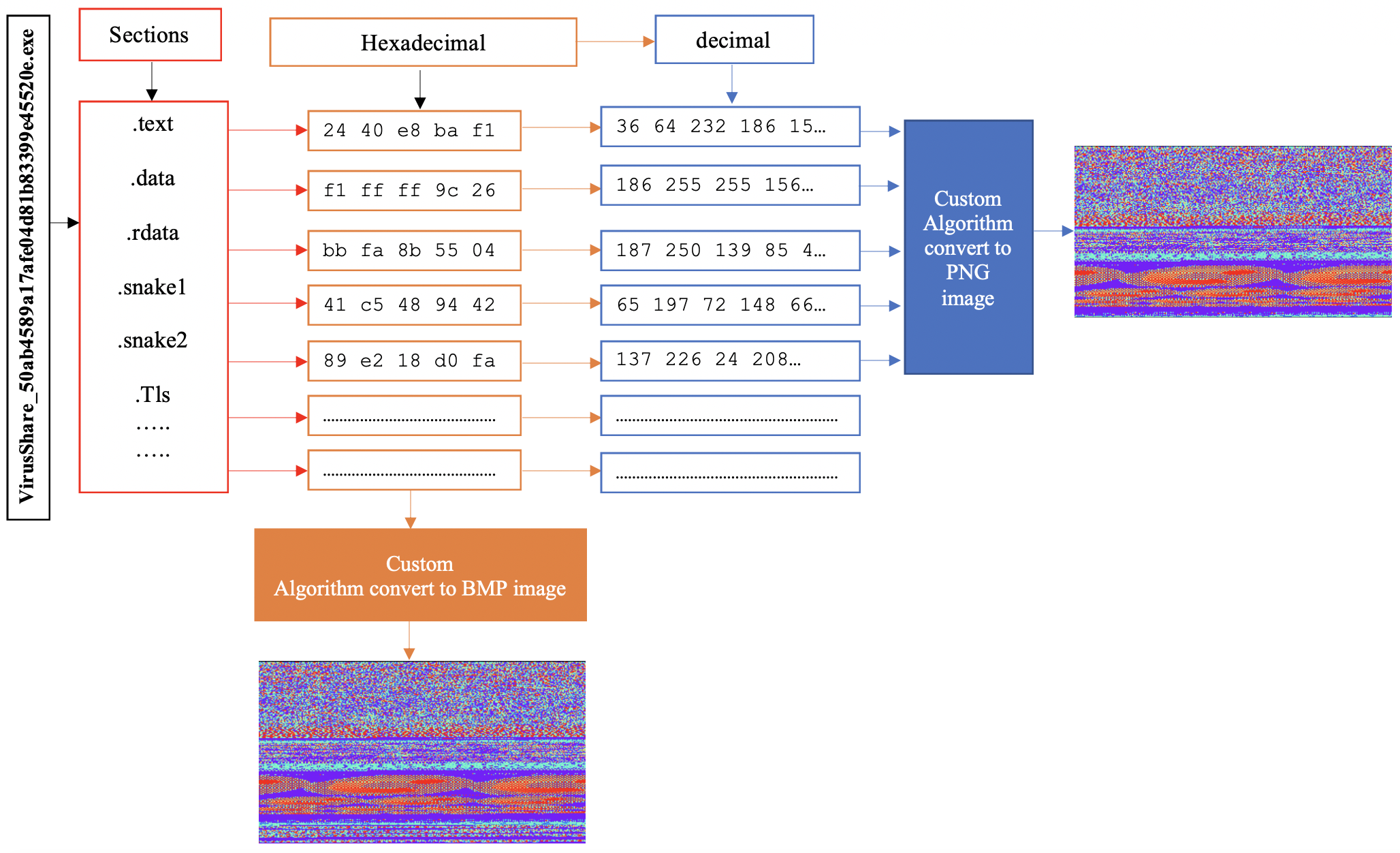}
	\caption{Converting malware to BMP and PNG Image}
\end{figure}

\begin{figure} [hbt]
	\includegraphics[width=\linewidth, frame]{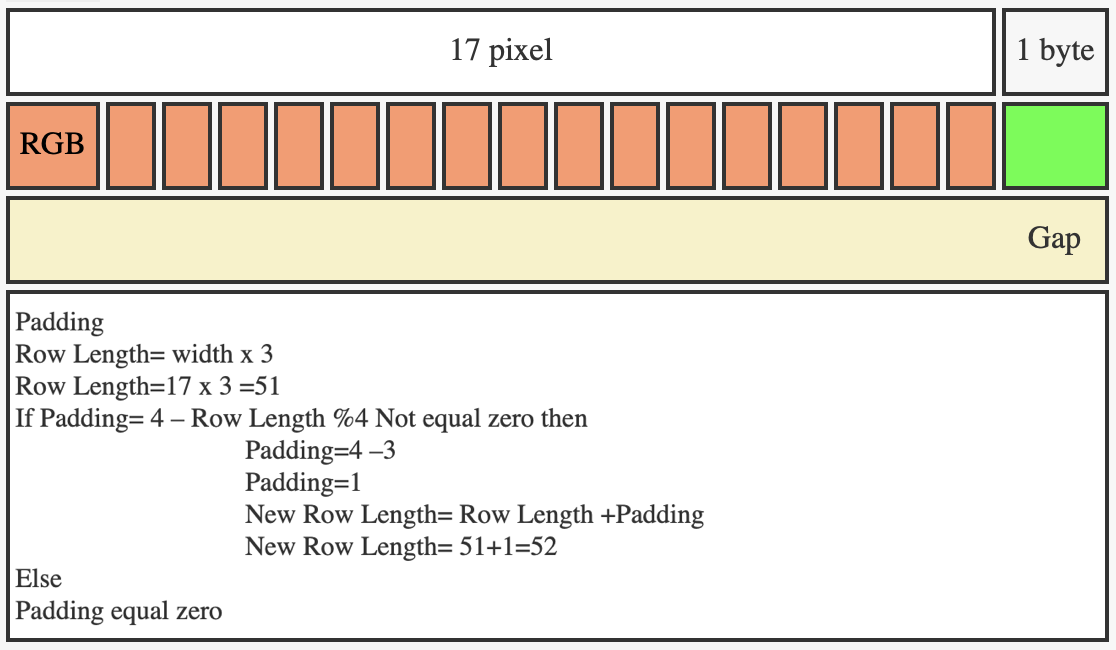}
	\caption{Padding Gap}
\end{figure}
\begin{figure} [hbt]
	\includegraphics[width=\linewidth, frame]{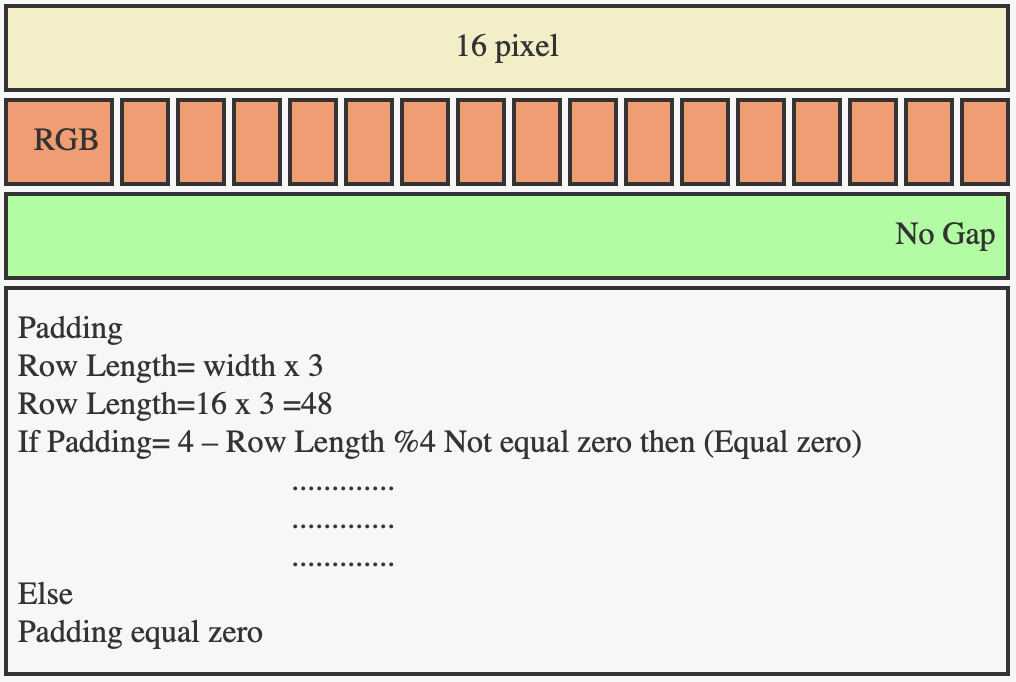}
	\caption{No Padding Gap}
\end{figure}

\section{Generative Adversarial Networks}

Generative Adversarial Networks (GANs) \cite{goodfellow2014generative} recently have achieved impressive results in image generation \cite{karras2019style, brock2018large, zhang2018photographic}. A GAN consists of a couple of neural networks that compete with each other. One is called the generator and the other the discriminator. The generator network tries to generate realistic samples that have never been seen before. The discriminator network tries to tell whether its inputs are real or fake. A GAN can be used to generate fake malware images. Specifically, it will improve the learning efficiency of our model by increasing the number of examples on what to train. 

Many applications of GANs in computer vision have achieved impressive performance such as SRGAN \cite{ledig2017photo}, pix2pix \cite{isola2017image}, CycleGAN \cite{zhu2017unpaired}, DiscoGAN \cite{ kim2017learning}, DualGAN \cite{yi2017dualgan} and StarGAN \cite{choi2018stargan}. Among these, CycleGAN is quite proficient in generating adversarial examples. It has achieved state-of-the-art results in image-to-image translation. CycleGAN has been applied to generate artificial training data if sufficient amount of real training data is not available. We use CycleGAN since we do not have enough Mac OS malware samples. Cycle-GAN is used to translate malware to benign images while simultaneously supervising an inverse benign to malware transformation model. 
\subsection{CycleGAN}
We use CycleGAN to map benign$(x)$ to malware$(y)$, and also map back to benign as shown in Fig 12. The CycleGAN model is used to convert Mac OS malware file to image using two mappings $G_{XY} : X \mapsto Y $ and $ G_{YX} : Y \mapsto X$, which fulfill the following constraints:\\
\begin{itemize}
\item 
\textbf{\emph{Generator}} $G:X \mapsto Y$ : translates images from  $X$ to  $Y$ (malware to benign)
\item 
\textbf{\emph{Generator}}  $F:Y \mapsto X$ : translates images from  $Y$ to  $X$ (benign to malware)
\item 
\textbf{\emph{Discriminator}} $DX$ : scores how real an image $X$ looks (Does this image look like benign?)
\item 
\textbf{\emph{Discriminator}} $DY$ : scores how real an image $Y$ looks like (Does this image look like a malware?)
\end{itemize}

\subsubsection{Objective function of CycleGAN}
As described above, the model relies on improving both the generator and discriminator.
We have two losses, one is adversarial Loss corresponding to the GAN and the second is Cycle Consistency Loss which computes how close the reconstructed image is to the original image.
\begin{itemize}
\item \textbf{\emph{Adversarial Loss}}

Adversarial loss is applied to both our mappings of generators and discriminators. The generator attempts to minimize the adversarial loss, and the discriminator attempts to maximize it. This adversarial loss shown below.
\begin{equation} \label{eq1}
\begin{split}
\mathcal {L}_{(G,D_y,X,Y)}= \mathbb{E}_{y\sim{p}_{data(y)}}[log (D_y(y))]+\\
\mathbb {E}_{x\sim{p}_{data(x)}}[log (1 - D_y(G(x)))]
\end{split}
\end{equation}

The equation above is to compute $D_x$ loss where ${p}_{data(y)}$ is the set of all malware image samples (represents the data distributions of y).
\begin{equation} \label{eq1}
\begin{split}
\mathcal{L}_{(G,D_x,Y,X)} =\mathbb{E}_{x\sim{p}_{data(x)}}[log(D_x(G(x)))]+\\
\mathbb{E}_{y\sim{p}_{data(y)}}[log(1 - D_x((G(y)))]
\end{split}
\end{equation}

The equation above is to compute $D_Y$ loss where ${p}_{data(x)}$ is the set of all benign image samples (represents the data distributions of x),\\

\item \textbf{\emph{Cycle Consistency Loss}}

CycleGan employs a forward and backward cycle consistency loss. The cycle consistency loss generates the content of the image efficiently while the image is translating. Cycle Consistency is computed as the difference between real and reconstructed images. In addition, two-cycle consistency losses guarantee that image that is transferred from one domain A to domain B, and back again will be the same. The formulation of cycle consistency is shown below:\\
\begin{equation} \label{eq1}
\begin{split}
\mathcal{L}_{(G,F)} =\mathbb{E}_{x\sim {p}_{data(x)}}[\Vert F(G(x))-x\Vert 1]+\\
\mathbb{E}_{y\sim {p}_{data(y)}}[\Vert F(G(y))-y\Vert 1]
\end{split}
\end{equation}

where,
$\Vert F(G(x))-x\Vert 1$ is the forward cycle consistency loss, and $\Vert F(G(y))-y\Vert 1$ is the backward cycle consistency loss\\

\end{itemize}
\subsubsection{Combined Objective Function}

The combined objective function is the summation of adversarial loss when converting malware to benign and benign to malware and their corresponding cycle consistency losses. 

$\mathcal{L}_{(G,F,D_x,D_y)} =\mathcal{L}_{(G,D_y,Y,X)}+\mathcal{L}_{(G,D_x,X,Y)}+\mathcal{L}_{(G,F)}$

\subsubsection{Optimization}
CycleGan performs optimization corresponding to:\\ 
$\mathcal{L}_{(G,F,D_x,D_y)}$ to come up with the generator that generates malware from benign and benign from malware.

\[G^*,F^*=\arg  \hspace{0.2cm} \underset{G,F}{min} \hspace{0.2cm} \underset{D_x,D_y}{ max}\hspace{0.2cm}\mathcal{L}{(G,F,D_x,D_y)}\]
\section{\textbf{Our Multi-task Learning Model}}
Several malware image samples are large, more than 1 megabyte. Therefore, large-scale malware image classification is a challenging task. In fact, when we used the state-of-the-art CNN models such as InceptionV3 \cite{szegedy2015going}, VGG19 \cite{simonyan2014very}, and ResNet \cite{he2016deep} for malware image binary classification, the accuracy rate came out very low. The issue is that the computational cost for large-scale image classification becomes unacceptable. We want a model with a capacity to learn from 100,000 malware images. We also need high-performance computing systems with GPUs or TPUs for classifying large public malware image repositories.
MTL can be used to save both time and memory needed by a learning system. How can we build an architecture for multi-task learning for large-scale malware images? To answer this question, we first create multiple deep CNNs for deep multi-task learning to develop joint training for different combination of tasks. Second, we design a network that can take input with different sizes of malware images.  In particular, the width and height for input malware images are determined by the malware file size. 
We designed a MTL model consisting of seven  classification tasks for malware image classification. As shown in Fig. 13, our deep multi-task learning algorithm is developed for joint training of multiple deep CNNs. Each CNN model contains 5 convolutional layers with PReLU activation function. The first four convolutional layers are followed by a max-pooling layers with a stride of 2 and the fifth convolutional layer is followed by two fully connected (FC) layers with PReLU activation function. Each FC layer consists of 1024 neurons as shown in Fig. 14.

\subsection{Convolutional neural network (CNN) layers}
A CNN is able to scale up to hundreds of layers with improving performance. In our model, we created large 9x9 filter for the first layer to shrink a large malware image to a moderate size. Besides, when we designed the model, we kept adding layers until the model over-fit at layer 12. Therefore, we designed the model with 11 layers. In addition, the feature map size should be large at the beginning and then a decrease. It is an acceptable assumption since a large malware image needs high resolution feature maps.

\begin{figure*} [ht]
	\includegraphics[width=\linewidth, frame]{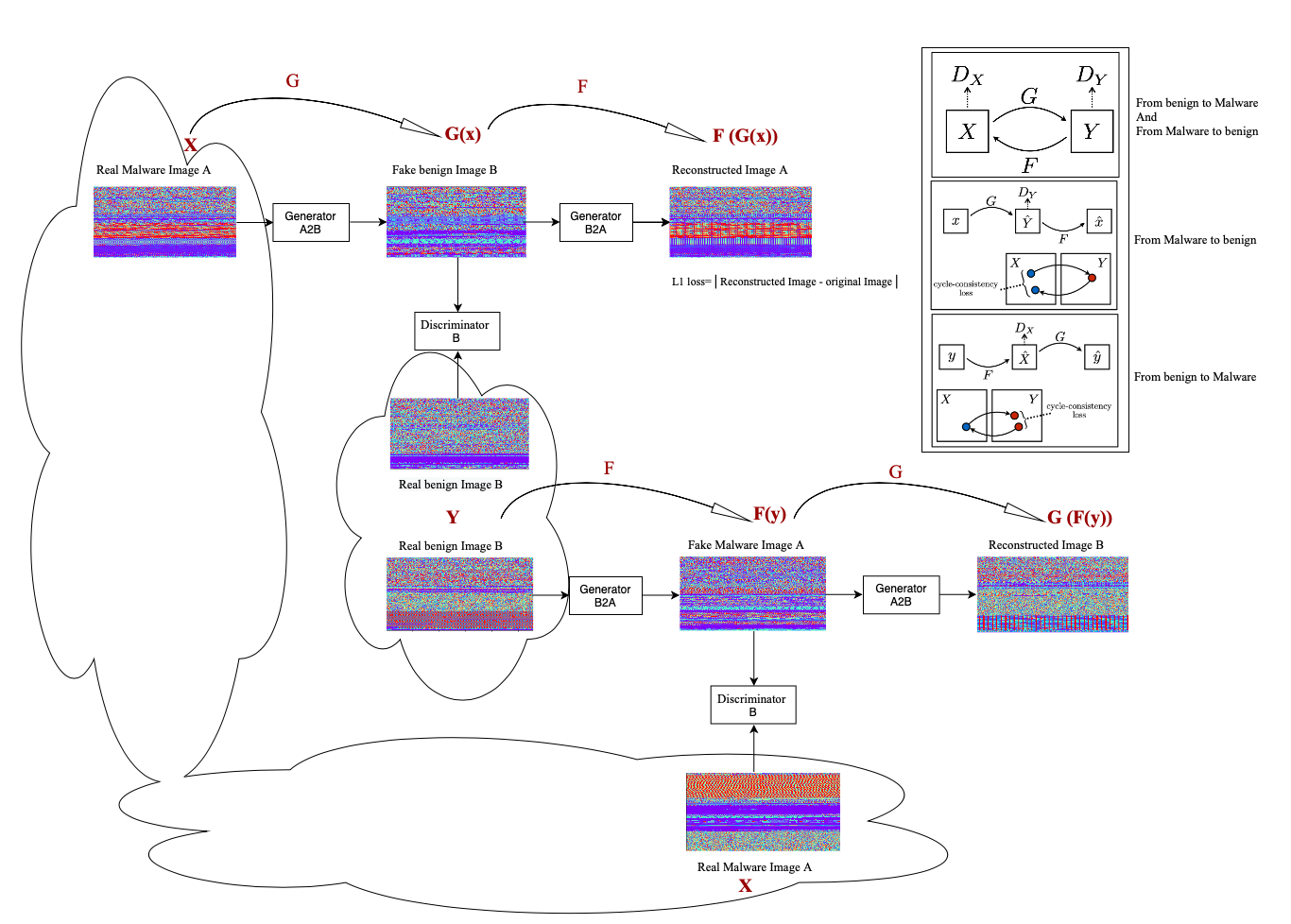}
	\caption{Generating malware images using CycleGAN}
\end{figure*}

\begin{figure*} [ht]
	\includegraphics[width=16cm, height=13cm, frame]{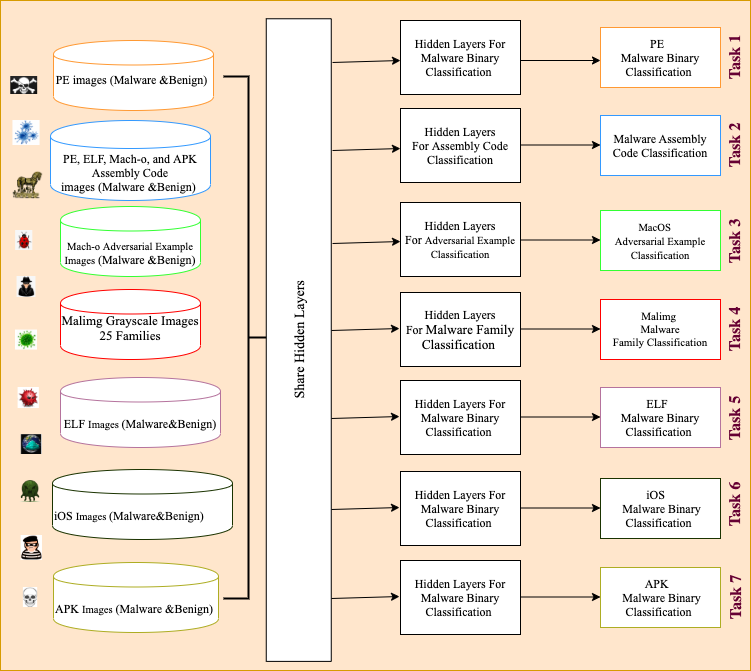}
	\caption{Our multi-task learning model}
\end{figure*}

\begin{figure*} [ht]
\centering
	\includegraphics[width=\linewidth]{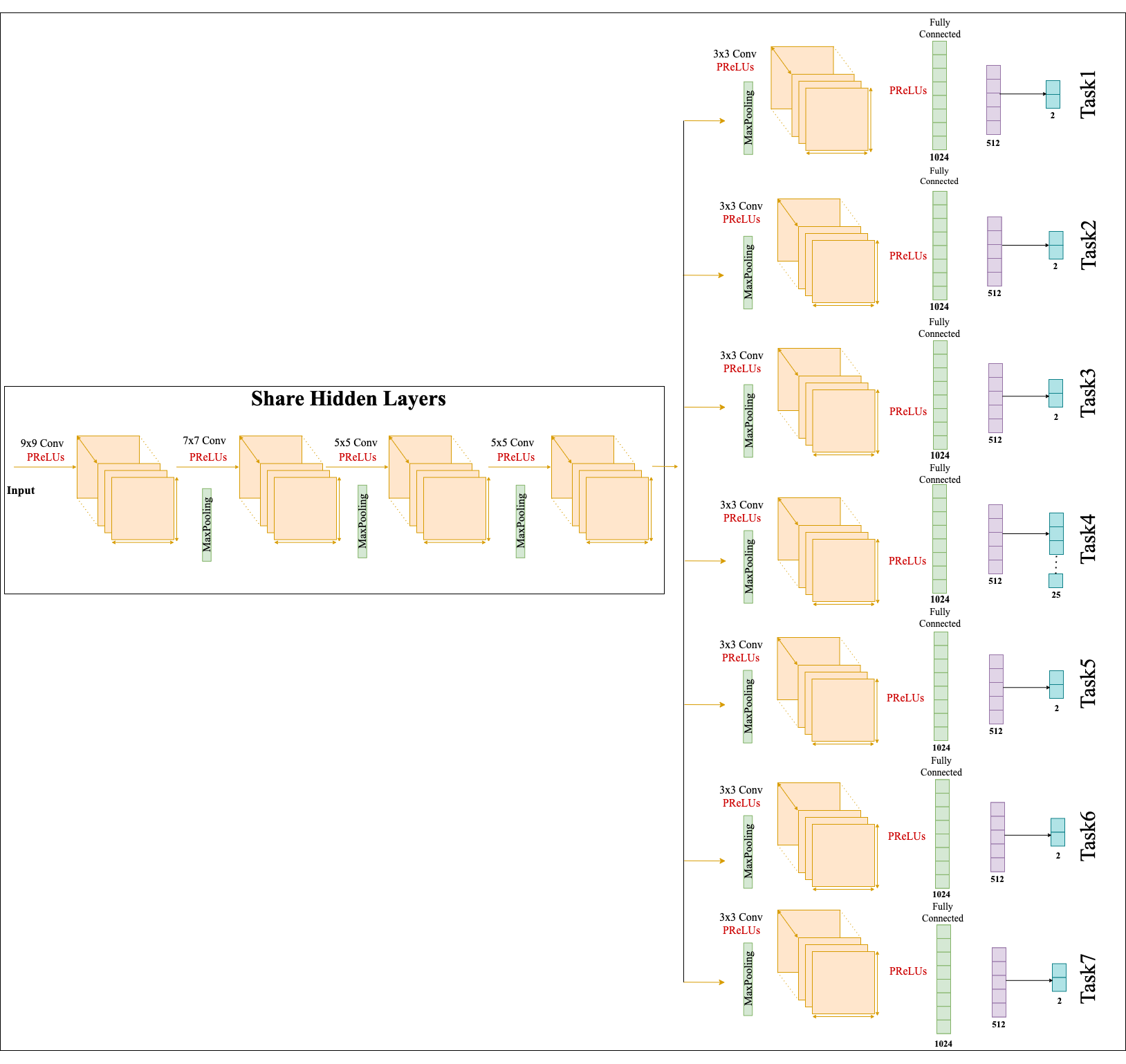}
	\caption{Our MTL model in detail}
\end{figure*}

\subsection{Activations}
An activation function maps a node's inputs to its output. We use four different activation functions that lead to different levels of performance in our model.

\subsubsection{Rectified Linear Unit (ReLU)} 
ReLU is the default activation function in deep learning. We use ReLU since it allows faster training time. It has been successfully applied in various state-of-the-art deep neural networks \cite{gao2019pixel, huang2017densely}. It is defined as $f(x)=max (0,x)$, where $x$ is the input of the activation
function \cite{nair2010rectified}. In other words,

$f(x) = \left \{ \begin{array}{rcl}
0 & \mbox{for} & x < 0,\\ x & \mbox{for} & x \ge 0.\end{array} \right.$\\

\subsubsection{LeakyReLU}
LeakyReLU replaces the negative part of the ReLU with a linear function by using fixed negative slope of 0.01 \cite{maas2013rectifier}. LeakyReLU is
defined as:\\

$f(x) = \left \{ \begin{array}{rcl}
0.01 x & \mbox{for} & x < 0,\\ x & \mbox{for} & x \ge 0.\end{array} \right.$ \\

We experimented with LeakyReLU since it has been used in many Kaggle competition projects and achieved high performance \cite{mastromichalakis2020alrelu}.

\subsubsection{Parametric Rectified Linear Units (PReLUs)}

\citet{he2015delving} proposed the parametric linear rectified (PReLU) to improve model fitting with nearly zero extra computational cost and little overfitting risk. In PReLU, the slope of negative part is learned from data instead of designing it as zero in ReLU. PReLU also generalizes the fixed small number used in LeakyReLU. PReLUs was shown to perform better than ReLU in a large scale image classification task \cite{wang2020alcoholism}. The function is defined as:\\

$f(x) = \left \{ \begin{array}{rcl}
a_i x & \mbox{for} & x \leq 0,\\ x & \mbox{for} & x > 0.\end{array} \right.$ \\

\noindent where $i$ is the index of a channel and $a_i$ is a learnable parameter. 
\subsubsection{Exponential linear unit (ELU)}
The Exponential Linear Unit (ELU) is based on ReLU. It reduces the bias shift effect, and pushes the negative inputs to be close to zero \cite{clevert2015fast}. It also speeds up learning, leading to better classification accuracy. ELU is
defined as:  \\

$f(x) = \left \{ \begin{array}{rcl} \alpha (exp(x) - 1) & \mbox{for} & x \le 0,\\ x & \mbox{for} & x > 0.\end{array} \right.$\\

We used ELU in our experiments since it has achieved best published result on CIFAR-10 and CIFAR-100 \cite{alom2020improved}

\subsection{Optimization}
An optimizer is used in training to minimize the loss and to make our prediction as accurate as possible. Most deep learning models use stochastic gradient descent  (SGD) \cite{ruder2016overview}. 
Many adaptive variants of SGD have been invented, including Adam \cite{ kingma2014adam}, Adagrad \cite{duchi2011adaptive}, Adadelta \cite{ zeiler2012adadelta}, RMSprop \cite{loizou2020momentum}, and Nadam \cite{dozat2016incorporating}. These five optimizers are used separately in our experiments, producing different results with our model as shown in Table 1. Among these adaptive optimizers, the Adam optimizer achieved the best performance. In summary, we compared four activation functions and five optimizers. The best performing activation function was PReLU and the best performing optimizer was Adam. On the other hand, the worst performing activation function was ReLU and the worst performing optimizer was Adadelta. 

\begin{table}[width=.4\textwidth,cols=2,pos=h]
\sffamily
  \caption{Comparing five optimizers on the model.}
\begin{tabular*}{\tblwidth}{@{}LL@{}}
\toprule
Optimizer                                           & Average Accuracy \\ 
\midrule
Adam \cite{kingma2014adam}         & 99.97\%          \\
Adagrad \cite{duchi2011adaptive}   & 64.36\%          \\
Adadelta \cite{zeiler2012adadelta} & 53.12\%          \\
RMSprop \cite{loizou2020momentum}  & 58.94\%          \\
RESprop \cite{loizou2020momentum}  & 60.43\%          \\ 
\bottomrule
\end{tabular*}
\end{table}

\section{Datasets}

We built several benchmark color Bitmap and JPG image datasets for Windows, Android, Linux, MacOS, and iOS operating systems of various malware executable files. The dataset samples were collected from Virushare by BitTorrent\footnote{\url{https://www.bittorrent.com}} that contains APKs, ELFs, EXEs, and DLLs from May 2019 to Sep 2020. Mach-o and IPAs samples were collected from VirusTotal and Contagio websites from Jun 2018 to Nov 2020.

Seven malware datasets were used in the experiments. The first one is a PE malware and benign color image dataset. The second dataset is the Mach-o Adversarial example malware and benign color image dataset. The third dataset is the ELF malware and benign color image dataset. The fourth dataset is Assembly code from PE, ELF, Mach-o, and APK malware and benign image dataset. The fifth dataset is a APK malware and benign color image dataset. The sixth dataset is a Mach-o and iOS malware and benign color image dataset. The seventh dataset is the Malimg grayscale image dataset \cite{nataraj2011malware} which contains 9339 malware images belonging to 25 different malware families. This dataset is publicly available.

We show a set of images obtained from malicious PE, Android, ELF, Mac, iOS samples. For instance, Fig. 15 shows the images for 8 different malicious samples.

\begin{figure}
	\includegraphics[width=\linewidth, frame]{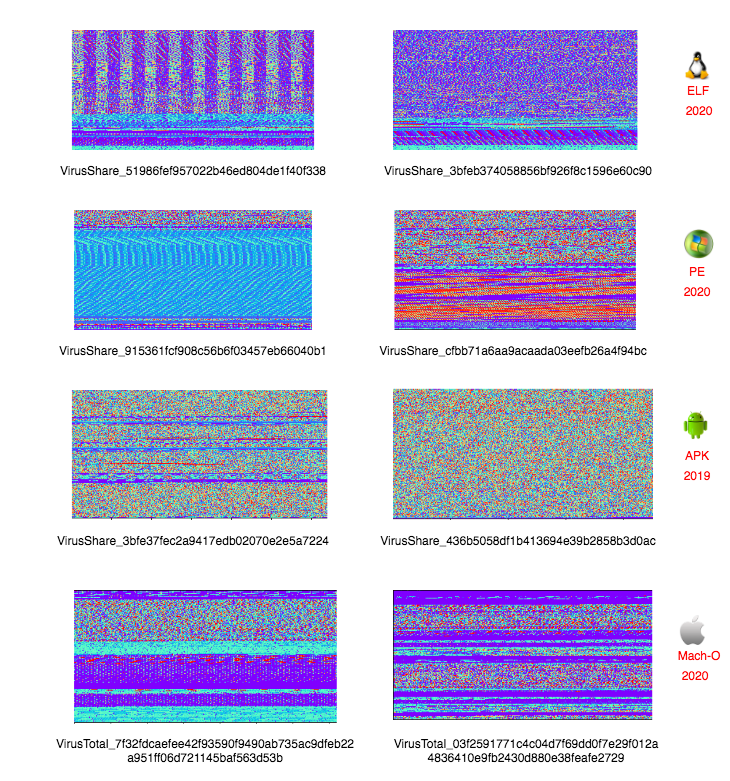}
	\caption{Malware samples from different operating systems}
		\label{FIG:1}
\end{figure}


\section{Evaluation}
We evaluate the our model's capability by using the following performance metrics: True Positive Rate (TPR), False Positive Rate (FPR) and Accuracy. TPR is the rate of malware samples correctly classified. FPR is the rate of malware samples falsely classified. \\

$TPR(Recall)=\frac{TP}{TP+FN}$\\

$Precision=\frac{TP}{TP+FP}$\\


$Accuracy=\frac{TP+TN}{TP+FN+FP+TN}$\\

$F\mbox{-}measure=\frac{2\times(Recall\times Precision)}{(Precision+Recall)}$\\

$Error Rate=\frac{(FP+FN)}{(TP+TN+FP+FN)}$\\

We repeated testing our model 5 times, and got the results as they are shown in Table 4. The accuracy of our model was between $99.80\% - 99.97\%$, and the average accuracy was $99.91\%$, with the highest accuracy of $99.97\%$. 

\begin{figure}
    \centering
    \subfloat[\centering Confusion Matrix for task1]{{\includegraphics[width=4cm, height=4cm]{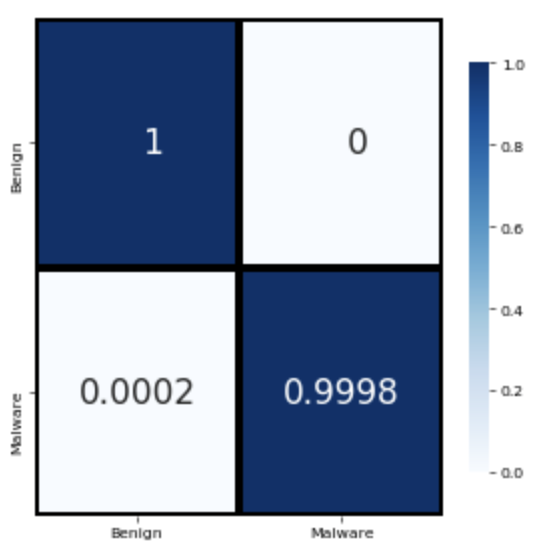} }}%
    \subfloat[\centering Confusion Matrix for task2]{{\includegraphics[width=4cm,height=4cm]{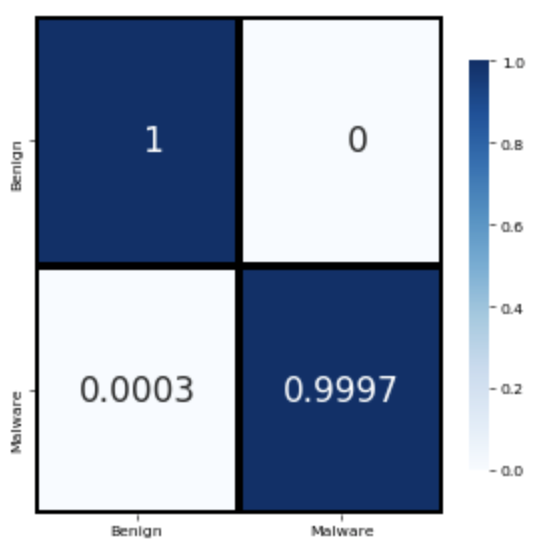} }}%
     \qquad
    \subfloat[\centering Confusion Matrix for task3]{{\includegraphics[width=4cm,height=4cm]{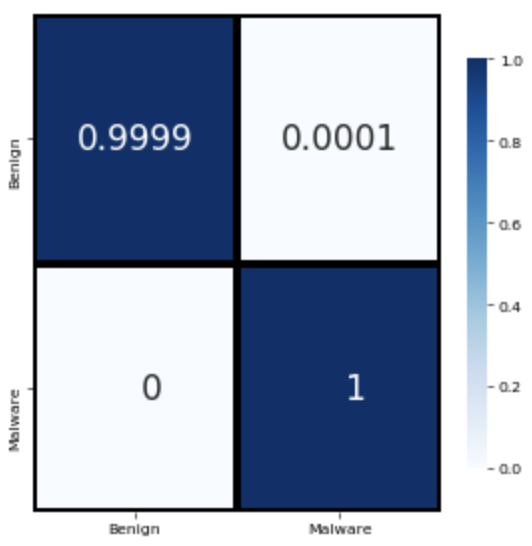} }}%
    \subfloat[\centering Confusion Matrix for task5]{{\includegraphics[width=4cm,height=4cm]{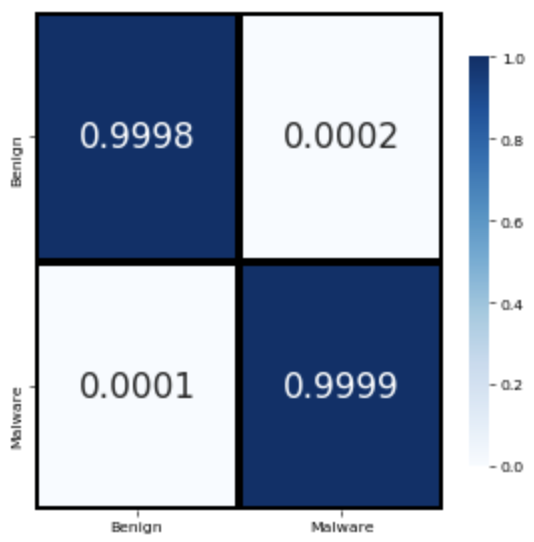} }}%
    \qquad
    \subfloat[\centering Confusion Matrix for task6]{{\includegraphics[width=4cm,height=4cm]{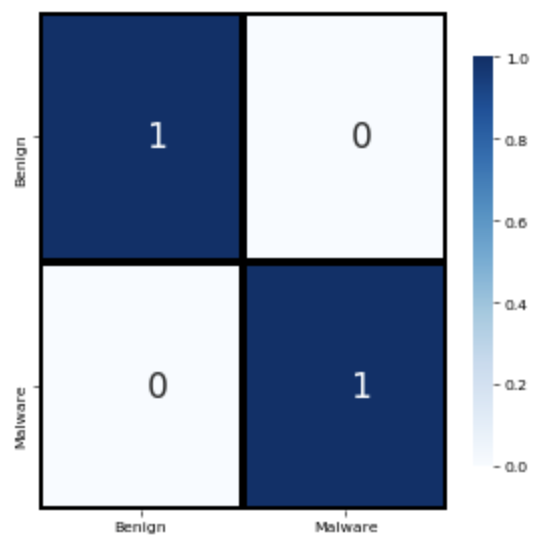} }}%
    \subfloat[\centering Confusion Matrix for task7]{{\includegraphics[width=4cm,height=4cm]{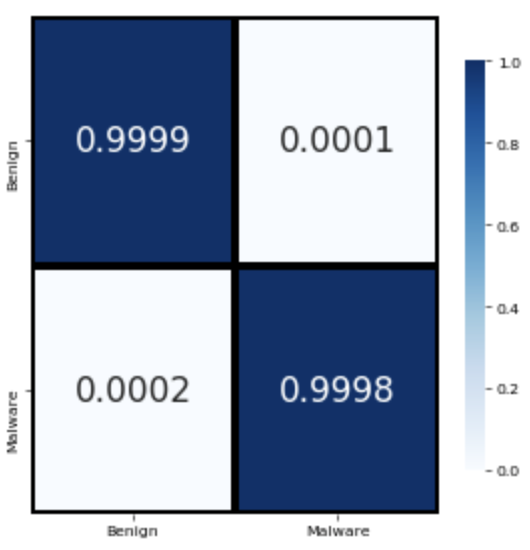} }}%

    \caption{Confusion Matrix for malware image binary classification}%
    \label{fig:example}%
\end{figure}

\begin{figure*} [h!]
\centering
	\includegraphics[width=\linewidth, frame]{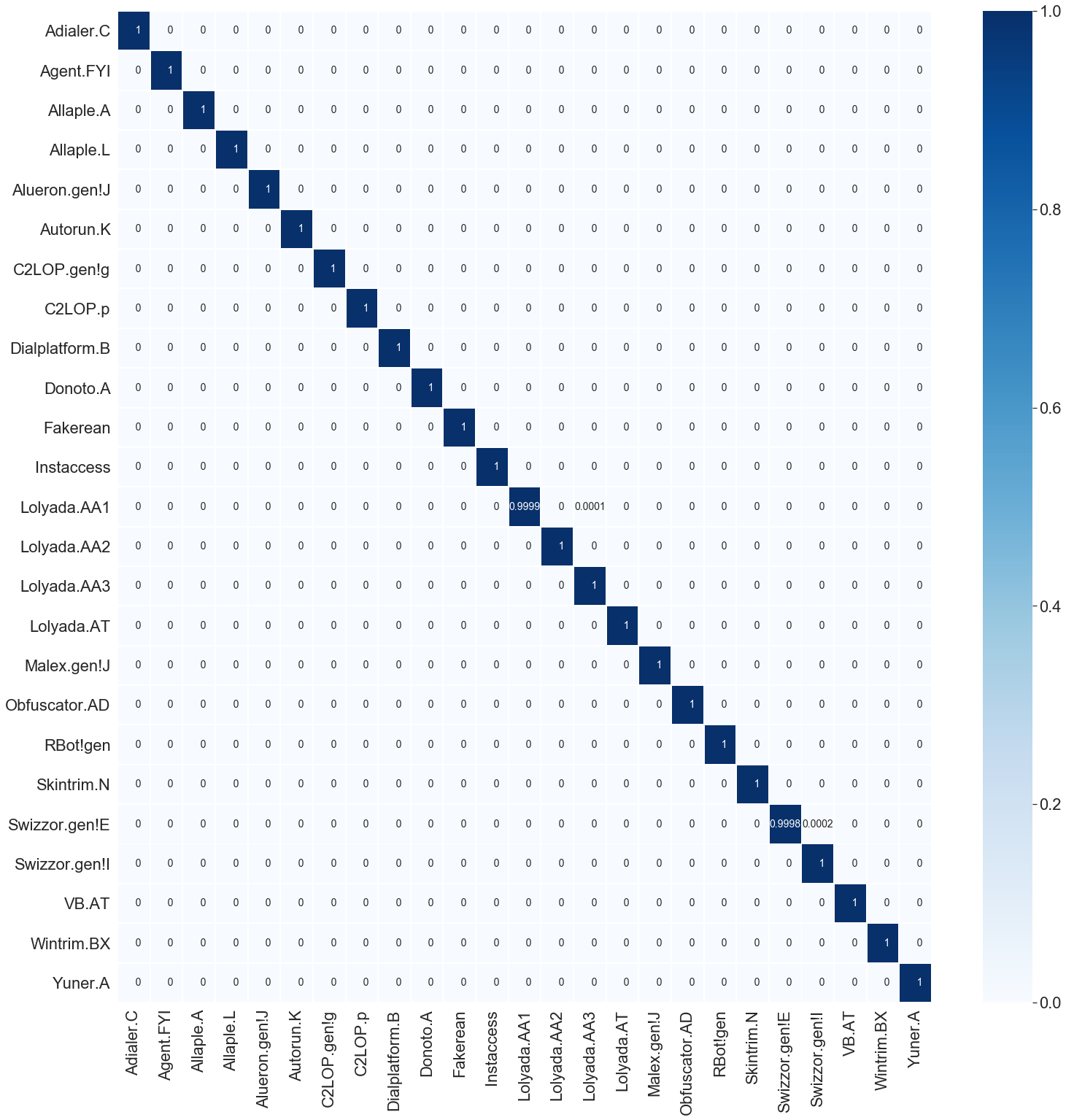}
	\caption{Confusion matrix for malware family classification-task4}
\end{figure*}
Fig. 16 and Fig. 17 show the confusion matrix for all tasks for predicting the classes of the examples in the testing dataset. All confusion matrices show superior performance on the diagonal. In Fig. 17, in the confusion matrix for task4, we observe that there is little misclassification and negligible errors, which means that classes close to the diagonal line have almost no similarities with other samples. As exception, the Lolyada.AA1 malware family has similarities with Lolyada.AA3 and Swizzor.gen!l with Swizzor.gen!E as well. These malware families have a little similarity behaviors. 

We also test our model for binary classification of malware versus benign in task1, task2, task3, task5, task6, and task 7. The confusion matrices for all these tasks are given in Figure 16.
\section{Experimental Setup}

We implemented the model using TensorFlow. The experiments were conducted on Google Colaboratory (also known as Colab). Google Colab is an execution environment that allows developers to write, run, and share code within Google Drive. Our datasets were mounted to the Google Colab using Google Drive. We used Python 3 with deep learning libraries and the model trained for about 8 hours. 
Our experimental results are illustrated in Table2 and Table3. One can see that our deep multi-task learning method has achieved competitive results.
\subsection{\textbf{Experimental Results}}
We evaluated our model performance and compared it with several other models. We show that multi-task learning can give a significant decrease in classification error. The training errors of four activation functions were measured. Fig. 18 shows the testing error rates using our datasets. Fig. 19 shows that accuracy of PReLU is better than the other functions for training. In Fig. 20, the accuracies for PReLU, LeakReLU, and ELU are 99.97\%, 99.91\%, and 98.57\%, respectively, while ReLU achieves 93.61\% accuracy. We also compared the performance of the model with the existing deep learning approaches based on testing accuracy using the Malimg benchmark dataset, as shown in Table 2. 
Among the four activation functions, the PReLU had the lowest training error rate of 0.0003 and lowest test error rate of 0.0005 while ReLU had the highest training error rate 0.0017, and highest test error rate 0.0027 as shown in Table 3. The model performed best when using the PReLU activation function. Fig. 17 shows the testing error rates for the four activation functions. The PReLU activation function is faster than the other activation functions with the lowest error rate. Fig. 18 shows the average accuracy of each activation function on all seven tasks. For the malware image binary classification task1, we obtain an accuracy of 99.88\%, 99.94\% for task2, 99.91\% for task3, 99.89\% for task5, 99.92\% for task6, and 99.95\% for task7, while task4 for the malware family classification yields a classification accuracy of 99.97\% as shown in Table 5. Fig. 16 shows the confusion matrix for malware image binary classification for task1, task2, task3, task5, task6, and task7.

We conclude that the use of PReLU gives better performance than using other activation functions. In addition, most research on malware image classification converts malware into gray scale images. We demonstrate that color images are more effective in malware image classification. Our study also has shown that using multi-task learning on malware images achieves the currently highest accuracy rate in malware image classification.

In addition, after trained combination all seven tasks $\{\tau_1,\tau_2,\tau_3,\tau_4,\tau_5,\tau_6,\tau_7\}$ we tested 900 samples of malware obfuscation that use various techniques on selected individual tasks  $\{\tau_1\},\{\tau_2\},\{\tau_3\},\{\tau_4\}\{\tau_5\},\{\tau_6\},\{\tau_7\}$ they effectively detect all malware obfuscated techniques. Therefore, there is no impact of malware obfuscation on our model. Table 6 shows the number of samples for each malware obfuscation techniques that we tested. 
In table 5 we randomly choose multiple different tasks for testing together to see how to improve the testing accuracy and comparing with testing accuracy of each single task. For example, the accuracy of $\{\tau_1,\tau_2\}$ is more than $\{\tau_1\}$ and $\{\tau_2\}$ separately, also the accuracy of $\{\tau_1,\tau_2,\tau_3,\tau_4\}$
is more than $\{\tau_1,\tau_2,\tau_4\}$. We find that training more task together is improve the accuracy. However, experimental evaluation on seven tasks malware image classification trained by the MTL improves learning performance rather than single-task learning (STL) approach by ~3.32\%, as shown in table 5. For comparison between our model and the state-of-the-art approaches is given in Table 7. Note that Task 3 and Task 6 are shown that no comparison with the state-of-the-art since the iOS and MacOS malware classification uncovered research using images.  

\begin{figure} [ht]
	\includegraphics[width=\linewidth, frame]{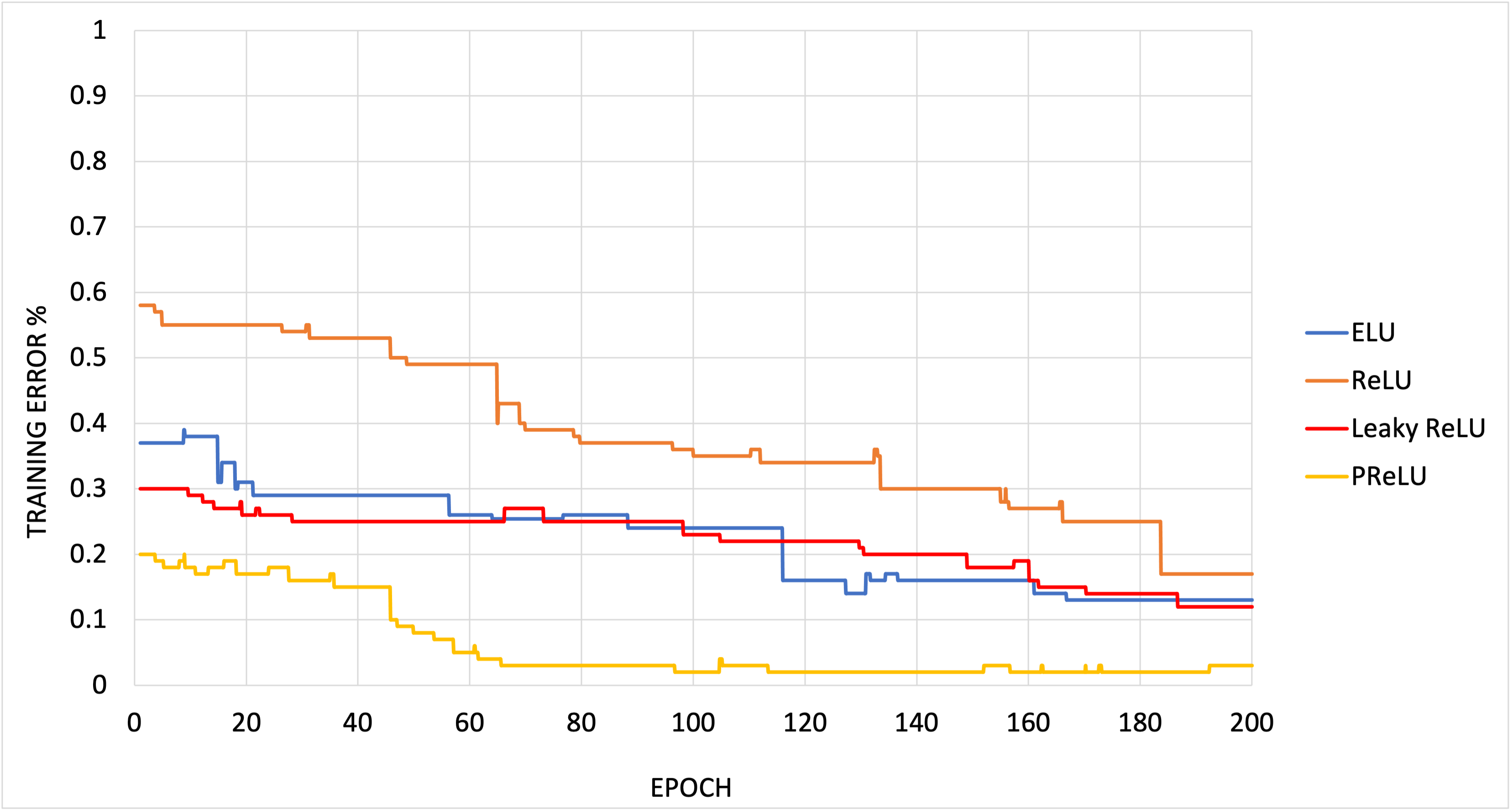}
	\caption{Testing error rates for four activation function}
\end{figure}

\begin{figure} [ht]
\centering
	\includegraphics[width=\linewidth, frame]{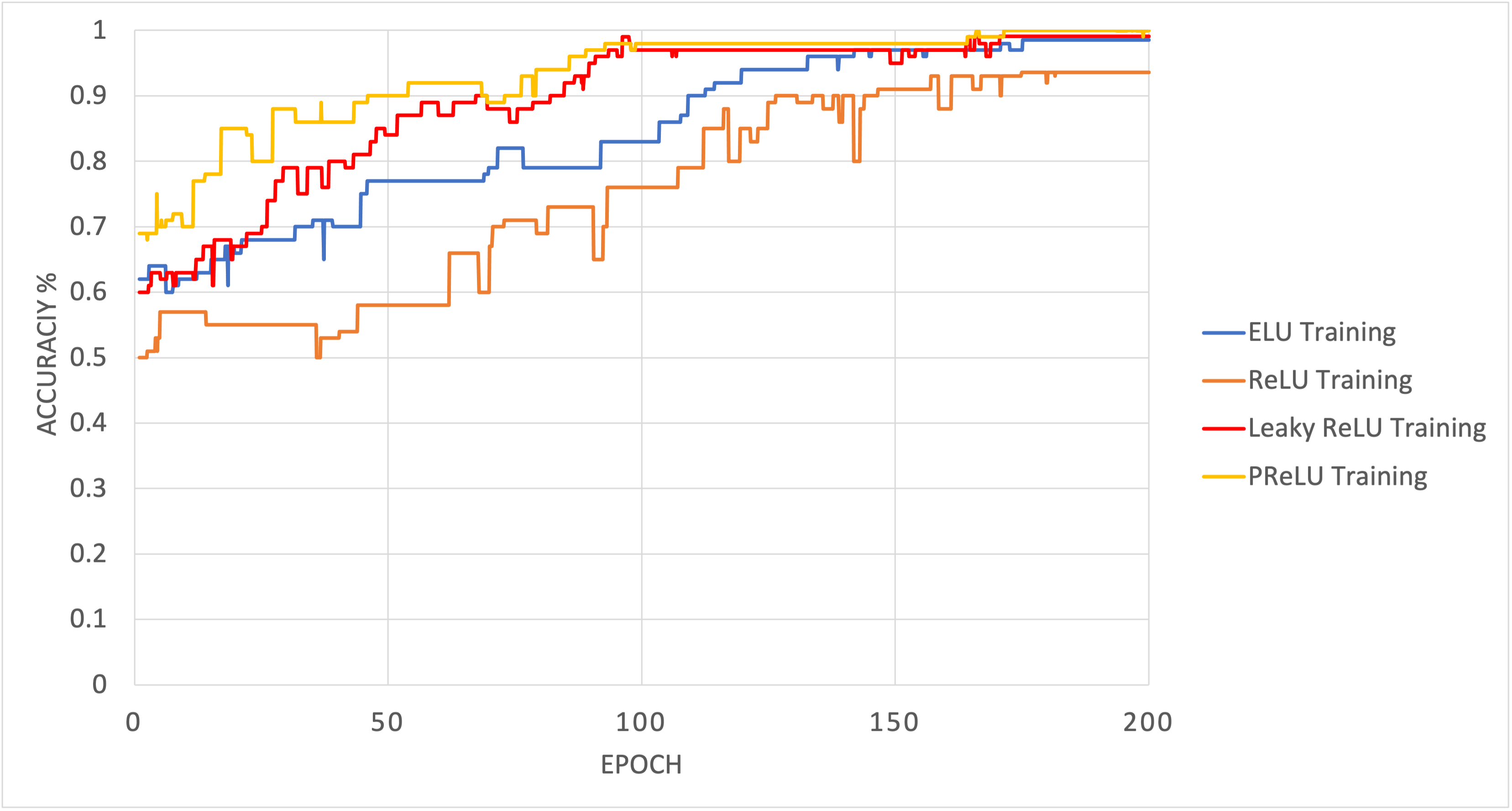}
	\caption{The accuracy during training for each activation function}
\end{figure}

\begin{figure} [ht]
\centering
	\includegraphics[width=\linewidth, frame]{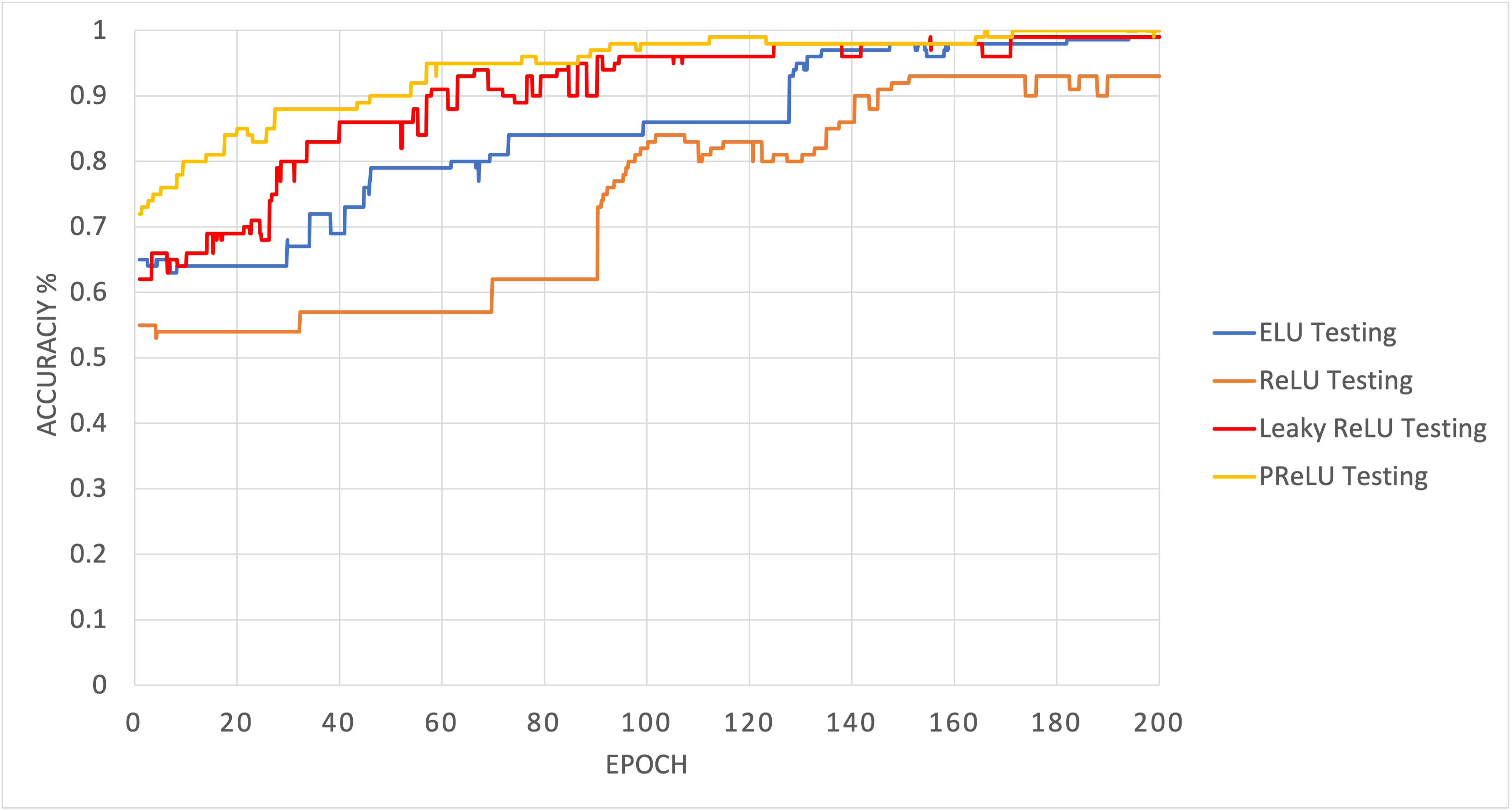}
	\caption{The accuracy during testing for each activation function}
\end{figure}

\begin{figure} [ht]
\centering
	\includegraphics[width=\linewidth, frame]{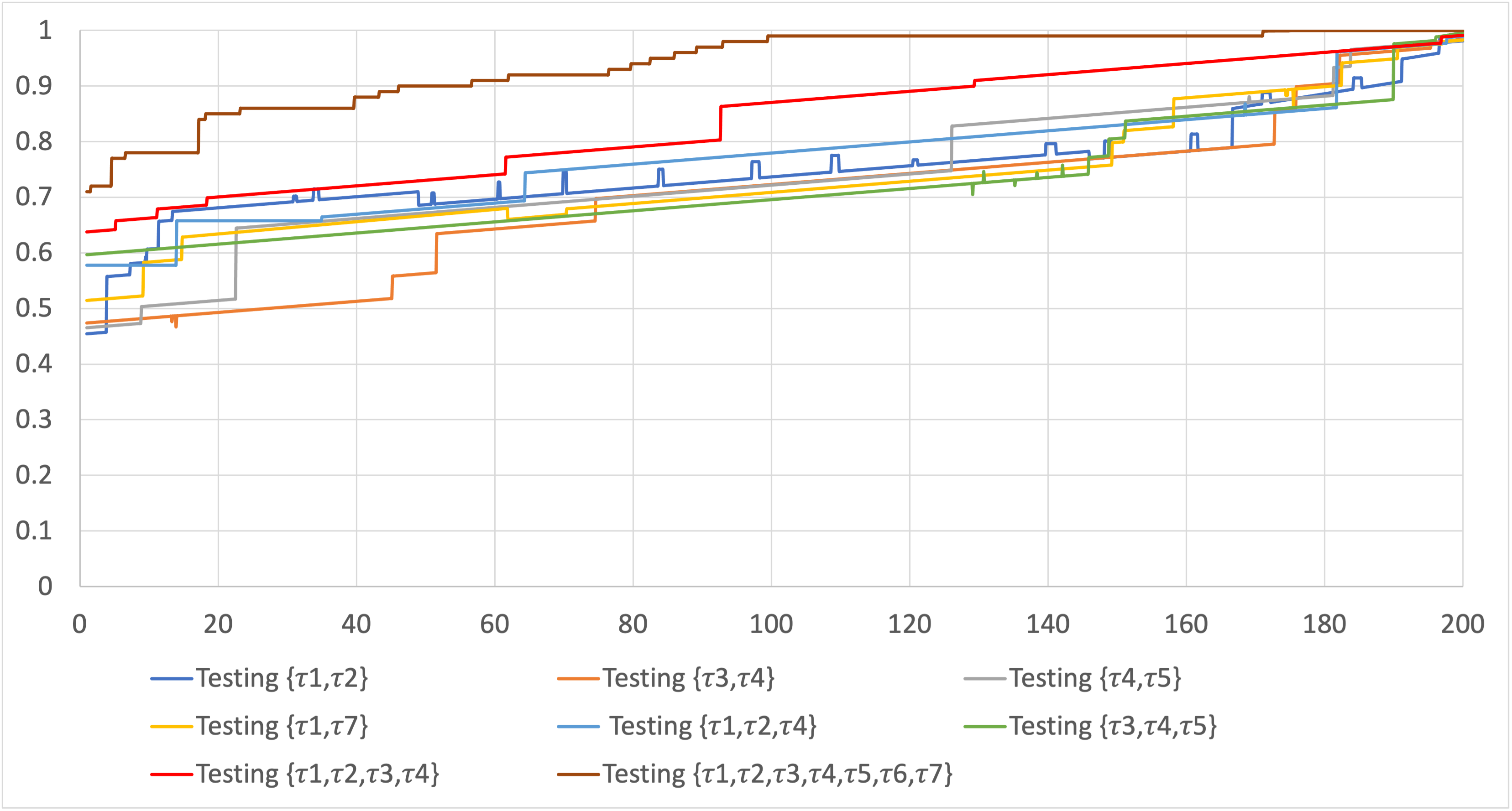}
	\caption{Testing accuracy on different tasks}
\end{figure}

\begin{table*}[width=.9\textwidth,cols=4,pos=h]
  \caption{Comparison between the average accuracy of the proposed model and some state-of-the-art models.}
  \begin{tabular*}{\tblwidth}{@{} LLLL@{} }
   \toprule
    Method & Dataset images & Files & Accuracy\\
   \midrule
IDA+DRBA \cite{cui2018detection}             & Malimg                        & PE                      & 94.50\%           \\ 
GIST + KNN \cite{cui2018detection}           & Malimg                        & PE                      & 91.90\%           \\ 
GIST + SVM \cite{cui2018detection}           & Malimg                        & PE                      & 92.20\%           \\ 
GLCM + KNN \cite{cui2018detection}           & Malimg                        & PE                      & 92.50\%           \\ 
GLCM + SVM \cite{cui2018detection}           & Malimg                        & PE                      & 93.20\%           \\ 
VGG16 + Softmax \cite{cui2018detection}                             & Malimg                        & PE                      & 90.77\%           \\ 
VGG16 + SVM \cite{rezende2018malicious}      & Malimg                        & PE                      & 92.97\%           \\ 
Light-weight DL \cite{su2018lightweight}     & Malimg                        & PE                      & 94.00\%           \\ 
NSGA-II \cite{cui2019malicious}              & Malimg                        & PE                      & 97.60\%           \\ 
ResNet + Sofrmax \cite{rezende2017malicious} & Malimg                        & PE                      & 98.62\%           \\ 
IMCEC \cite{vasan2020image}                  & Malimg                        & PE                      & 99.50\%           \\ 
IMCFN \cite{vasan2020imcfn}                  & RGB Malimg                    & PE                      & 98.82\%           \\ 
Vgg-verydeep-19 \cite{yue2017imbalanced}     & Malimg                        & PE                      & 97.18\%           \\ 
Custom CNN \cite{gibert2019using}            & Malimg                        & PE                      & 98.48\%           \\ 
Custom CNN \cite{gibert2019using}            & Microsoft malware             & PE                      & 97.49\%           \\ 
VisDroid \cite{bakour2020visdroid}                        & Android Drebin                & APK                     & 98.14\%           \\ 
Random Forest \cite{yang2017detecting}                    & Android Drebin                & APK                     & 95.42\%             \\ 
Inception V3  \cite{bensaoud2020classifying}                                                  & Malimg                        & PE                      & 99.24\%           \\ 
Our multi-task learning model                                   & Malimg + Malware color images & PE + ELF + APK + Mach-o & \textbf{99.97\%}  \\ 
   \bottomrule
  \end{tabular*}
\end{table*}

\begin{table}[width=.45\textwidth,cols=3,pos=h]
\sffamily
  \caption{Error rate for four different activation functions.}
  \begin{tabular*}{\tblwidth}{@{} LLL@{} }
   \toprule
    Activation & Training Error & Test Error\\
   \midrule
ReLU       & 0.0017        & 0.0027     \\
Leaky ReLU & 0.0012        & 0.0022     \\ 
ELU        & 0.0013        & 0.0021     \\ 
PReLU      & 0.0003        & 0.0005     \\ 
   \bottomrule
  \end{tabular*}
\end{table}

\begin{table}[width=.45 \textwidth,cols=4,pos=h]
\sffamily
  \caption{Testing the model 5 times.}
  \begin{tabular*}{\tblwidth}{@{} LLLL@{} }
   \toprule
    No & Accuracy (\%) & TPR (\%) & FPR (\%)\\
   \midrule
1 & 99.87 & 99.88 & \textbf{0.45} \\ 
2 & 99.90 & 99.92 & 0.50 \\ 
3 & 99.91 & 99.93 & 0.65 \\ 
4 & 99.94 & 99.95 & 0.66 \\ 
5 & \textbf{99.97} & \textbf{99.98} & 0.73 \\ 
   \bottomrule
  \end{tabular*}
\end{table}


\begin{table*}[width=.99\textwidth,cols=8,pos=h]
\caption{Comparison the testing among all tasks results.}
\begin{tabular}{@{}llllllll@{}}
\toprule
Tasks                           & \{$\tau_1$\}   & \{$\tau_2$\}   & \{$\tau_3$\}   & \{$\tau_4$\}  & \{$\tau_5$\} & \{$\tau_6$\}  & \{$\tau_7$\}   \\ \midrule
\{$\tau_1$\}                         & 96.56\% &       &       &       &       &       &       \\
\{$\tau_2$\}                         &       & 97.02\% &       &       &       &       &       \\
\{$\tau_3$\}                         &       &       & 96.89\% &       &       &       &       \\
\{$\tau_4$\}                         &       &       &       & 97.86\% &       &       &       \\
\{$\tau_5$\}                         &       &       &       &       & 96.67\% &       &       \\
\{$\tau_6$\}                         &       &       &       &       &       & 96.11\% &       \\
\{$\tau_7$\}                         &       &       &       &       &       &       & 96.84\% \\
\{$\tau_1,\tau_2$\}                     & 98.36\% & 97.56\% &       &       &       &       &       \\
\{$\tau_3,\tau_4$\}                     &       &       & 98.12\% & 98.72\% &       &       &       \\
\{$\tau_4,\tau_5$\}                     &       &       &       & 98.80\% & 98.45\% &       &       \\
\{$\tau_1,\tau_7$\}                     & 98.02\% &       &       &       &       &       & 98.39\% \\
\{$\tau_1,\tau_2,\tau_4$\}                 & 99.10\% & 99.12\% &       & 99.03\% &       &       &       \\
\{$\tau_3,\tau_4,\tau_5$\}                 &       &       & 99.05\% & 99.24\% & 99.15\% &       &       \\
\{$\tau_1,\tau_2,\tau_3,\tau_4$\}             & 99.19\% & 99.22\% & 99.20\% & 99.17\% &       &       &       \\
\{$\tau_1,\tau_2,\tau_3,\tau_4,\tau_5,\tau_6,\tau_7$\} & \textcolor{red}{\textbf{99.88\%}} & \textcolor{red}{\textbf{99.94\%}} & \textcolor{red}{\textbf{99.91\%}} & \textcolor{red}{\textbf{99.97\%}} & \textcolor{red}{\textbf{99.89\%}} & \textcolor{red}{\textbf{99.92\%}} & \textcolor{red}{\textbf{99.95\%}} \\ \bottomrule
\end{tabular}
\end{table*}


\begin{table}[width=4\textwidth,cols=3,pos=h]
\sffamily
\caption{Malware Obfuscation techniques.}
\begin{tabular}{@{}lll@{}}
\toprule
\# & Obfuscation Techniques                   & Samples \\ \midrule
1  & Disassembling \& Reassembling            & 50                \\
2  & Changing Package Name                    & 100               \\
3  & Repacking                                & 60                \\
4  & Identifier Renaming                      & 40                \\
5  & Call Indirections                        & 55                \\
6  & Junk Code Insertion                      & 30                \\
7  & Garbage Code Insertion                   & 100               \\
8  & Register Usage Exchange                  & 70                \\
9  & Instruction replacement                  & 35                \\
10 & Dead code insertion                      & 40                \\
11 & Register reassignment                    & 60                \\
12 & Subroutine Permutation                   & 50                \\
13 & Code transposition                       & 70                \\
14 & Code Reordering through Jump Instruction & 60                \\
15 & Code integration (Zmist)                 & 80                \\ \bottomrule
\end{tabular}
\end{table}

\begin{table}[width=.48\textwidth,cols=4,pos=h]
\sffamily
  \caption{Comparing with the state-of-the-art methods for each task.}
\begin{tabular}{@{}
l 
l 
l 
c @{}}
\toprule
Tasks & Acuracy & Models                                        & Acuracy \\ \midrule
$\tau_1$   & 99.88\%   & MSIC \cite{Azab9104999}     & 96\%                                                  \\
$\tau_2$    & 99.94\%   & \citet{khan2019analysis}     & 88.36\%                                               \\
$\tau_3$    & 99.91\%   & \multicolumn{1}{c}{-} & -                                                   \\
$\tau_4$    & 99.97\%   & IMCFN \cite{vasan2020imcfn}  & 98.27\%                                               \\
$\tau_5$    & 99.89\%   & \citet{Su8377943}   & 94\%                                                   \\
$\tau_6$    & 99.92\%   & \multicolumn{1}{c}{-} & -                                                   \\
$\tau_6$    & 99.95\%   & IMCFN \cite{vasan2020imcfn}  & 97.35\%                                               \\ \bottomrule
\end{tabular}
\end{table}

\section{\textbf{Conclusions and Future Work}}

We proposed a novel method in this paper to detect malware by building a multi-task learning model and creating malware images. First, we generated images from various malware files for several operating systems: Android, Windows, MacOS, iOS, and Linux.  
We provided experimental results showing that our model can learn flexible shared parameters for related tasks, resulting in significantly improved performance over the state-of-the-art in malware image classification.  In addition, our experiments show that the PReLU activation function has better generalization performance than ELU, ReLU, and LeakyReLU in our model with 11 layers. Training for 11 layers required 175 epochs to give best accuracy for all tasks. If we have than 11 layers the model needs more than 250 epochs. \\
The focus of future work is to use unrelated tasks like assembly code text and  API calls text with our model to see if it hurts or improves performance of our model. 
\newpage
\bibliographystyle{model1-num-names.bst}
\medskip
\bibliography{cas-refs}

\begin{thebibliography}{71}
\expandafter\ifx\csname natexlab\endcsname\relax\def\natexlab#1{#1}\fi
\providecommand{\url}[1]{\texttt{#1}}
\providecommand{\href}[2]{#2}
\providecommand{\path}[1]{#1}
\providecommand{\DOIprefix}{doi:}
\providecommand{\ArXivprefix}{arXiv:}
\providecommand{\URLprefix}{URL: }
\providecommand{\Pubmedprefix}{pmid:}
\providecommand{\doi}[1]{\href{http://dx.doi.org/#1}{\path{#1}}}
\providecommand{\Pubmed}[1]{\href{pmid:#1}{\path{#1}}}
\providecommand{\bibinfo}[2]{#2}
\ifx\xfnm\relax \def\xfnm[#1]{\unskip,\space#1}\fi
\bibitem[{McAfee(2020)}]{report}
\bibinfo{author}{J.~D. McAfee}, \bibinfo{title}{{Malware Threats Statistics}},
  \bibinfo{year}{2020}.
\bibitem[{Ucci et~al.(2019)Ucci, Aniello, and Baldoni}]{ucci2019survey}
\bibinfo{author}{D.~Ucci}, \bibinfo{author}{L.~Aniello},
  \bibinfo{author}{R.~Baldoni},
\newblock \bibinfo{title}{Survey of machine learning techniques for malware
  analysis},
\newblock \bibinfo{journal}{Computers \& Security} \bibinfo{volume}{81}
  (\bibinfo{year}{2019}) \bibinfo{pages}{123--147}.
\bibitem[{Alzaylaee et~al.(2020)Alzaylaee, Yerima, and Sezer}]{alzaylaee2020dl}
\bibinfo{author}{M.~K. Alzaylaee}, \bibinfo{author}{S.~Y. Yerima},
  \bibinfo{author}{S.~Sezer},
\newblock \bibinfo{title}{Dl-droid: Deep learning based android malware
  detection using real devices},
\newblock \bibinfo{journal}{Computers \& Security} \bibinfo{volume}{89}
  (\bibinfo{year}{2020}) \bibinfo{pages}{101663}.
\bibitem[{Darabian et~al.(2020)Darabian, Homayounoot, Dehghantanha, Hashemi,
  Karimipour, Parizi, and Choo}]{darabian2020detecting}
\bibinfo{author}{H.~Darabian}, \bibinfo{author}{S.~Homayounoot},
  \bibinfo{author}{A.~Dehghantanha}, \bibinfo{author}{S.~Hashemi},
  \bibinfo{author}{H.~Karimipour}, \bibinfo{author}{R.~M. Parizi},
  \bibinfo{author}{K.-K.~R. Choo},
\newblock \bibinfo{title}{Detecting cryptomining malware: a deep learning
  approach for static and dynamic analysis},
\newblock \bibinfo{journal}{Journal of Grid Computing}  (\bibinfo{year}{2020})
  \bibinfo{pages}{1--11}.
\bibitem[{Souri et~al.(2018)Souri, Navimipour, and Rahmani}]{souri2018formal}
\bibinfo{author}{A.~Souri}, \bibinfo{author}{N.~J. Navimipour},
  \bibinfo{author}{A.~M. Rahmani},
\newblock \bibinfo{title}{Formal verification approaches and standards in the
  cloud computing: a comprehensive and systematic review},
\newblock \bibinfo{journal}{Computer Standards \& Interfaces}
  \bibinfo{volume}{58} (\bibinfo{year}{2018}) \bibinfo{pages}{1--22}.
\bibitem[{Sihag et~al.(2021)Sihag, Vardhan, and Singh}]{SIHAG2021100365}
\bibinfo{author}{V.~Sihag}, \bibinfo{author}{M.~Vardhan},
  \bibinfo{author}{P.~Singh},
\newblock \bibinfo{title}{A survey of android application and malware
  hardening},
\newblock \bibinfo{journal}{Computer Science Review} \bibinfo{volume}{39}
  (\bibinfo{year}{2021}) \bibinfo{pages}{100365}.
\bibitem[{Qiu et~al.(2020)Qiu, Zhang, Luo, Pan, Nepal, and
  Xiang}]{qiu2020survey}
\bibinfo{author}{J.~Qiu}, \bibinfo{author}{J.~Zhang}, \bibinfo{author}{W.~Luo},
  \bibinfo{author}{L.~Pan}, \bibinfo{author}{S.~Nepal},
  \bibinfo{author}{Y.~Xiang},
\newblock \bibinfo{title}{A survey of android malware detection with deep
  neural models},
\newblock \bibinfo{journal}{ACM Computing Surveys (CSUR)} \bibinfo{volume}{53}
  (\bibinfo{year}{2020}) \bibinfo{pages}{1--36}.
\bibitem[{Ngo et~al.(2020)Ngo, Nguyen, Le, and Nguyen}]{NGO2020280}
\bibinfo{author}{Q.-D. Ngo}, \bibinfo{author}{H.-T. Nguyen},
  \bibinfo{author}{V.-H. Le}, \bibinfo{author}{D.-H. Nguyen},
\newblock \bibinfo{title}{A survey of iot malware and detection methods based
  on static features},
\newblock \bibinfo{journal}{ICT Express} \bibinfo{volume}{6}
  (\bibinfo{year}{2020}) \bibinfo{pages}{280--286}.
\bibitem[{Nataraj et~al.(2011)Nataraj, Karthikeyan, Jacob, and
  Manjunath}]{nataraj2011malware}
\bibinfo{author}{L.~Nataraj}, \bibinfo{author}{S.~Karthikeyan},
  \bibinfo{author}{G.~Jacob}, \bibinfo{author}{B.~S. Manjunath},
  \bibinfo{title}{{Malware Images: Visualization and Automatic
  classification}}, \bibinfo{year}{2011}.
\bibitem[{Vasan et~al.(2020)Vasan, Alazab, Wassan, Safaei, and
  Zheng}]{vasan2020image}
\bibinfo{author}{D.~Vasan}, \bibinfo{author}{M.~Alazab},
  \bibinfo{author}{S.~Wassan}, \bibinfo{author}{B.~Safaei},
  \bibinfo{author}{Q.~Zheng},
\newblock \bibinfo{title}{{Image-Based Malware Classification Using Ensemble of
  CNN Architectures (IMCEC)}},
\newblock \bibinfo{journal}{Computers \& Security}  (\bibinfo{year}{2020})
  \bibinfo{pages}{101748}.
\bibitem[{Su et~al.(2018)Su, Vasconcellos, Prasad, Daniele, Feng, and
  Sakurai}]{su2018lightweight}
\bibinfo{author}{J.~Su}, \bibinfo{author}{V.~D. Vasconcellos},
  \bibinfo{author}{S.~Prasad}, \bibinfo{author}{S.~Daniele},
  \bibinfo{author}{Y.~Feng}, \bibinfo{author}{K.~Sakurai},
  \bibinfo{title}{{Lightweight Classification of IoT Malware Based on Image
  Recognition}}, \bibinfo{year}{2018}.
\bibitem[{Ni et~al.(2018)Ni, Qian, and Zhang}]{ni2018malware}
\bibinfo{author}{S.~Ni}, \bibinfo{author}{Q.~Qian}, \bibinfo{author}{R.~Zhang},
\newblock \bibinfo{title}{{Malware Identification using Visualization Images
  and Deep Learning}},
\newblock \bibinfo{journal}{Computers \& Security} \bibinfo{volume}{77}
  (\bibinfo{year}{2018}) \bibinfo{pages}{871--885}.
\bibitem[{Bensaoud et~al.(2020)Bensaoud, Abudawaood, and
  Kalita}]{bensaoud2020classifying}
\bibinfo{author}{A.~Bensaoud}, \bibinfo{author}{N.~Abudawaood},
  \bibinfo{author}{J.~Kalita},
\newblock \bibinfo{title}{Classifying malware images with convolutional neural
  network models},
\newblock \bibinfo{journal}{International Journal of Network Security}
  \bibinfo{volume}{22} (\bibinfo{year}{2020}) \bibinfo{pages}{1022--1031}.
\bibitem[{Naeem et~al.(2020)Naeem, Ullah, Naeem, Khalid, Vasan, Jabbar, and
  Saeed}]{naeem2020malware}
\bibinfo{author}{H.~Naeem}, \bibinfo{author}{F.~Ullah}, \bibinfo{author}{M.~R.
  Naeem}, \bibinfo{author}{S.~Khalid}, \bibinfo{author}{D.~Vasan},
  \bibinfo{author}{S.~Jabbar}, \bibinfo{author}{S.~Saeed},
\newblock \bibinfo{title}{Malware detection in industrial internet of things
  based on hybrid image visualization and deep learning model},
\newblock \bibinfo{journal}{Ad Hoc Networks}  (\bibinfo{year}{2020})
  \bibinfo{pages}{102154}.
\bibitem[{Kalash et~al.(2018)Kalash, Rochan, Mohammed, Bruce, Wang, and
  Iqbal}]{kalash2018malware}
\bibinfo{author}{M.~Kalash}, \bibinfo{author}{M.~Rochan},
  \bibinfo{author}{N.~Mohammed}, \bibinfo{author}{N.~D. Bruce},
  \bibinfo{author}{Y.~Wang}, \bibinfo{author}{F.~Iqbal},
  \bibinfo{title}{Malware classification with deep convolutional neural
  networks}, \bibinfo{year}{2018}.
\bibitem[{Mercaldo and Santone(2020)}]{mercaldo2020deep}
\bibinfo{author}{F.~Mercaldo}, \bibinfo{author}{A.~Santone},
\newblock \bibinfo{title}{Deep learning for image-based mobile malware
  detection},
\newblock \bibinfo{journal}{Journal of Computer Virology and Hacking
  Techniques}  (\bibinfo{year}{2020}) \bibinfo{pages}{1--15}.
\bibitem[{Meyerson and Miikkulainen(2018)}]{meyerson2018pseudo}
\bibinfo{author}{E.~Meyerson}, \bibinfo{author}{R.~Miikkulainen},
\newblock \bibinfo{title}{Pseudo-task augmentation: From deep multitask
  learning to intratask sharing---and back},
\newblock \bibinfo{journal}{arXiv preprint arXiv:1803.04062}
  (\bibinfo{year}{2018}).
\bibitem[{Zhang et~al.(2020)Zhang, Zheng, Liu, and Jia}]{zhang2020deep}
\bibinfo{author}{K.~Zhang}, \bibinfo{author}{L.~Zheng},
  \bibinfo{author}{Z.~Liu}, \bibinfo{author}{N.~Jia},
\newblock \bibinfo{title}{A deep learning based multitask model for
  network-wide traffic speed prediction},
\newblock \bibinfo{journal}{Neurocomputing} \bibinfo{volume}{396}
  (\bibinfo{year}{2020}) \bibinfo{pages}{438--450}.
\bibitem[{Gkioxari et~al.(2014)Gkioxari, Hariharan, Girshick, and
  Malik}]{gkioxari2014r}
\bibinfo{author}{G.~Gkioxari}, \bibinfo{author}{B.~Hariharan},
  \bibinfo{author}{R.~Girshick}, \bibinfo{author}{J.~Malik},
\newblock \bibinfo{title}{{R-cnns for Pose Estimation and Action Detection}},
\newblock \bibinfo{journal}{arXiv preprint arXiv:1406.5212}
  (\bibinfo{year}{2014}).
\bibitem[{Zhao et~al.(2020)Zhao, Peng, and He}]{zhao2020attribute}
\bibinfo{author}{J.~Zhao}, \bibinfo{author}{Y.~Peng}, \bibinfo{author}{X.~He},
\newblock \bibinfo{title}{Attribute hierarchy based multi-task learning for
  fine-grained image classification},
\newblock \bibinfo{journal}{Neurocomputing} \bibinfo{volume}{395}
  (\bibinfo{year}{2020}) \bibinfo{pages}{150--159}.
\bibitem[{Wah et~al.(2011)Wah, Branson, Welinder, Perona, and
  Belongie}]{wah2011caltech}
\bibinfo{author}{C.~Wah}, \bibinfo{author}{S.~Branson},
  \bibinfo{author}{P.~Welinder}, \bibinfo{author}{P.~Perona},
  \bibinfo{author}{S.~Belongie},
\newblock \bibinfo{title}{The caltech-ucsd birds-200-2011 dataset}
  (\bibinfo{year}{2011}).
\bibitem[{Krause et~al.(2013)Krause, Stark, Deng, and Fei-Fei}]{krause20133d}
\bibinfo{author}{J.~Krause}, \bibinfo{author}{M.~Stark},
  \bibinfo{author}{J.~Deng}, \bibinfo{author}{L.~Fei-Fei},
\newblock \bibinfo{title}{3d object representations for fine-grained
  categorization},
\newblock in: \bibinfo{booktitle}{Proceedings of the IEEE international
  conference on computer vision workshops}, \bibinfo{year}{2013}, pp.
  \bibinfo{pages}{554--561}.
\bibitem[{Bell et~al.(2020)Bell, Liu, Alsheikh, Tang, Pizzi, Henning, Singh,
  Parkhi, and Borisyuk}]{bell2020groknet}
\bibinfo{author}{S.~Bell}, \bibinfo{author}{Y.~Liu},
  \bibinfo{author}{S.~Alsheikh}, \bibinfo{author}{Y.~Tang},
  \bibinfo{author}{E.~Pizzi}, \bibinfo{author}{M.~Henning},
  \bibinfo{author}{K.~Singh}, \bibinfo{author}{O.~Parkhi},
  \bibinfo{author}{F.~Borisyuk}, \bibinfo{title}{Groknet: Unified computer
  vision model trunk and embeddings for commerce}, \bibinfo{year}{2020}.
\bibitem[{Yu et~al.(2020)Yu, Chen, Wang, Xian, Chen, Liu, Madhavan, and
  Darrell}]{yu2020bdd100k}
\bibinfo{author}{F.~Yu}, \bibinfo{author}{H.~Chen}, \bibinfo{author}{X.~Wang},
  \bibinfo{author}{W.~Xian}, \bibinfo{author}{Y.~Chen},
  \bibinfo{author}{F.~Liu}, \bibinfo{author}{V.~Madhavan},
  \bibinfo{author}{T.~Darrell}, \bibinfo{title}{Bdd100k: A diverse driving
  dataset for heterogeneous multitask learning}, \bibinfo{year}{2020}.
\bibitem[{Chen et~al.(2020)Chen, Wang, Yu, Gao, Li, and Wang}]{chen2020spatial}
\bibinfo{author}{Z.~Chen}, \bibinfo{author}{R.~Wang}, \bibinfo{author}{M.~Yu},
  \bibinfo{author}{H.~Gao}, \bibinfo{author}{Q.~Li}, \bibinfo{author}{H.~Wang},
\newblock \bibinfo{title}{Spatial--temporal multi-task learning for salient
  region detection},
\newblock \bibinfo{journal}{Pattern Recognition Letters} \bibinfo{volume}{132}
  (\bibinfo{year}{2020}) \bibinfo{pages}{76--83}.
\bibitem[{Wang et~al.(2020)Wang, Chen, Ran, Luo, Chan, Tham, Chang, Mannil,
  Cheung, and Heng}]{wang2020towards}
\bibinfo{author}{X.~Wang}, \bibinfo{author}{H.~Chen}, \bibinfo{author}{A.-R.
  Ran}, \bibinfo{author}{L.~Luo}, \bibinfo{author}{P.~P. Chan},
  \bibinfo{author}{C.~C. Tham}, \bibinfo{author}{R.~T. Chang},
  \bibinfo{author}{S.~S. Mannil}, \bibinfo{author}{C.~Y. Cheung},
  \bibinfo{author}{P.-A. Heng},
\newblock \bibinfo{title}{Towards multi-center glaucoma oct image screening
  with semi-supervised joint structure and function multi-task learning},
\newblock \bibinfo{journal}{Medical Image Analysis} \bibinfo{volume}{63}
  (\bibinfo{year}{2020}) \bibinfo{pages}{101695}.
\bibitem[{Dorado-Moreno et~al.(2020)Dorado-Moreno, Navarin, Guti{\'e}rrez,
  Prieto, Sperduti, Salcedo-Sanz, and
  Herv{\'a}s-Mart{\'\i}nez}]{dorado2020multi}
\bibinfo{author}{M.~Dorado-Moreno}, \bibinfo{author}{N.~Navarin},
  \bibinfo{author}{P.~A. Guti{\'e}rrez}, \bibinfo{author}{L.~Prieto},
  \bibinfo{author}{A.~Sperduti}, \bibinfo{author}{S.~Salcedo-Sanz},
  \bibinfo{author}{C.~Herv{\'a}s-Mart{\'\i}nez},
\newblock \bibinfo{title}{Multi-task learning for the prediction of wind power
  ramp events with deep neural networks},
\newblock \bibinfo{journal}{Neural Networks} \bibinfo{volume}{123}
  (\bibinfo{year}{2020}) \bibinfo{pages}{401--411}.
\bibitem[{Drepper(2006)}]{drepper2006write}
\bibinfo{author}{U.~Drepper},
\newblock \bibinfo{title}{{How to Write Shared Libraries}},
\newblock \bibinfo{journal}{Retrieved Jul} \bibinfo{volume}{16}
  (\bibinfo{year}{2006}) \bibinfo{pages}{2009}.
\bibitem[{Choi et~al.(2011)Choi, Zhu, and Lee}]{choi2011detecting}
\bibinfo{author}{H.~Choi}, \bibinfo{author}{B.~B. Zhu},
  \bibinfo{author}{H.~Lee},
\newblock \bibinfo{title}{{Detecting Malicious Web Links and Identifying Their
  Attack Types.}},
\newblock \bibinfo{journal}{WebApps} \bibinfo{volume}{11}
  (\bibinfo{year}{2011}) \bibinfo{pages}{218}.
\bibitem[{AppNee(2020)}]{PEiD}
\bibinfo{author}{AppNee}, \bibinfo{title}{{PEiD - Popular PE Packer
  /Cryptor/Compiler Detector}}, \bibinfo{year}{2020}.
\bibitem[{Smart et~al.(2005)Smart, Csomor et~al.}]{smart2005cross}
\bibinfo{author}{J.~Smart}, \bibinfo{author}{S.~Csomor}, et~al.,
  \bibinfo{title}{{Cross-Platform GUI Programming with wxWidgets}},
  \bibinfo{publisher}{Prentice Hall Professional}, \bibinfo{year}{2005}.
\bibitem[{Goodfellow et~al.(2014)Goodfellow, Pouget-Abadie, Mirza, Xu,
  Warde-Farley, Ozair, Courville, and Bengio}]{goodfellow2014generative}
\bibinfo{author}{I.~Goodfellow}, \bibinfo{author}{J.~Pouget-Abadie},
  \bibinfo{author}{M.~Mirza}, \bibinfo{author}{B.~Xu},
  \bibinfo{author}{D.~Warde-Farley}, \bibinfo{author}{S.~Ozair},
  \bibinfo{author}{A.~Courville}, \bibinfo{author}{Y.~Bengio},
\newblock \bibinfo{title}{Generative adversarial nets},
\newblock in: \bibinfo{booktitle}{Advances in neural information processing
  systems}, \bibinfo{year}{2014}, pp. \bibinfo{pages}{2672--2680}.
\bibitem[{Karras et~al.(2019)Karras, Laine, and Aila}]{karras2019style}
\bibinfo{author}{T.~Karras}, \bibinfo{author}{S.~Laine},
  \bibinfo{author}{T.~Aila},
\newblock \bibinfo{title}{A style-based generator architecture for generative
  adversarial networks},
\newblock in: \bibinfo{booktitle}{Proceedings of the IEEE/CVF Conference on
  Computer Vision and Pattern Recognition}, \bibinfo{year}{2019}, pp.
  \bibinfo{pages}{4401--4410}.
\bibitem[{Brock et~al.(2018)Brock, Donahue, and Simonyan}]{brock2018large}
\bibinfo{author}{A.~Brock}, \bibinfo{author}{J.~Donahue},
  \bibinfo{author}{K.~Simonyan},
\newblock \bibinfo{title}{Large scale gan training for high fidelity natural
  image synthesis},
\newblock \bibinfo{journal}{arXiv preprint arXiv:1809.11096}
  (\bibinfo{year}{2018}).
\bibitem[{Zhang et~al.(2018)Zhang, Xie, and Yang}]{zhang2018photographic}
\bibinfo{author}{Z.~Zhang}, \bibinfo{author}{Y.~Xie},
  \bibinfo{author}{L.~Yang},
\newblock \bibinfo{title}{Photographic text-to-image synthesis with a
  hierarchically-nested adversarial network},
\newblock in: \bibinfo{booktitle}{Proceedings of the IEEE Conference on
  Computer Vision and Pattern Recognition}, \bibinfo{year}{2018}, pp.
  \bibinfo{pages}{6199--6208}.
\bibitem[{Ledig et~al.(2017)Ledig, Theis, Husz{\'a}r, Caballero, Cunningham,
  Acosta, Aitken, Tejani, Totz, Wang et~al.}]{ledig2017photo}
\bibinfo{author}{C.~Ledig}, \bibinfo{author}{L.~Theis},
  \bibinfo{author}{F.~Husz{\'a}r}, \bibinfo{author}{J.~Caballero},
  \bibinfo{author}{A.~Cunningham}, \bibinfo{author}{A.~Acosta},
  \bibinfo{author}{A.~Aitken}, \bibinfo{author}{A.~Tejani},
  \bibinfo{author}{J.~Totz}, \bibinfo{author}{Z.~Wang}, et~al.,
  \bibinfo{title}{{Photo-Realistic Single Image Super-Resolution Using a
  Generative Adversarial Network}}, \bibinfo{year}{2017}.
\bibitem[{Isola et~al.(????)Isola, Zhu, Zhou, and Efros}]{isola2017image}
\bibinfo{author}{P.~Isola}, \bibinfo{author}{J.-Y. Zhu},
  \bibinfo{author}{T.~Zhou}, \bibinfo{author}{A.~A. Efros},
  \bibinfo{title}{{Image-to-Image Translation with Conditional Adversarial
  Networks}}, ????
\bibitem[{Zhu et~al.(2017)Zhu, Park, Isola, and Efros}]{zhu2017unpaired}
\bibinfo{author}{J.-Y. Zhu}, \bibinfo{author}{T.~Park},
  \bibinfo{author}{P.~Isola}, \bibinfo{author}{A.~A. Efros},
  \bibinfo{title}{{Unpaired Image-to-Image Translation Using Cycle-Consistent
  Adversarial Networks}}, \bibinfo{year}{2017}.
\bibitem[{Kim et~al.(2017)Kim, Cha, Kim, Lee, and Kim}]{kim2017learning}
\bibinfo{author}{T.~Kim}, \bibinfo{author}{M.~Cha}, \bibinfo{author}{H.~Kim},
  \bibinfo{author}{J.~K. Lee}, \bibinfo{author}{J.~Kim},
\newblock \bibinfo{title}{{Learning to Discover Cross-Domain Relations with
  Generative Adversarial Networks}},
\newblock \bibinfo{journal}{arXiv preprint arXiv:1703.05192}
  (\bibinfo{year}{2017}).
\bibitem[{Yi et~al.(????)Yi, Zhang, Tan, and Gong}]{yi2017dualgan}
\bibinfo{author}{Z.~Yi}, \bibinfo{author}{H.~Zhang}, \bibinfo{author}{P.~Tan},
  \bibinfo{author}{M.~Gong}, \bibinfo{title}{{Dualgan: Unsupervised Dual
  Learning for Image-to-Image Translation}}, ????
\bibitem[{Choi et~al.(2018)Choi, Choi, Kim, Ha, Kim, and
  Choo}]{choi2018stargan}
\bibinfo{author}{Y.~Choi}, \bibinfo{author}{M.~Choi}, \bibinfo{author}{M.~Kim},
  \bibinfo{author}{J.-W. Ha}, \bibinfo{author}{S.~Kim},
  \bibinfo{author}{J.~Choo}, \bibinfo{title}{{Stargan: Unified Generative
  Adversarial Networks for Multi-Domain Image-to-Image translation}},
  \bibinfo{year}{2018}.
\bibitem[{Szegedy et~al.(2015)Szegedy, Liu, Jia, Sermanet, Reed, Anguelov,
  Erhan, Vanhoucke, and Rabinovich}]{szegedy2015going}
\bibinfo{author}{C.~Szegedy}, \bibinfo{author}{W.~Liu},
  \bibinfo{author}{Y.~Jia}, \bibinfo{author}{P.~Sermanet},
  \bibinfo{author}{S.~Reed}, \bibinfo{author}{D.~Anguelov},
  \bibinfo{author}{D.~Erhan}, \bibinfo{author}{V.~Vanhoucke},
  \bibinfo{author}{A.~Rabinovich},
\newblock \bibinfo{title}{Going deeper with convolutions},
\newblock in: \bibinfo{booktitle}{Proceedings of the IEEE conference on
  computer vision and pattern recognition}, \bibinfo{year}{2015}, pp.
  \bibinfo{pages}{1--9}.
\bibitem[{Simonyan and Zisserman(2014)}]{simonyan2014very}
\bibinfo{author}{K.~Simonyan}, \bibinfo{author}{A.~Zisserman},
\newblock \bibinfo{title}{Very deep convolutional networks for large-scale
  image recognition},
\newblock \bibinfo{journal}{arXiv preprint arXiv:1409.1556}
  (\bibinfo{year}{2014}).
\bibitem[{He et~al.(2016)He, Zhang, Ren, and Sun}]{he2016deep}
\bibinfo{author}{K.~He}, \bibinfo{author}{X.~Zhang}, \bibinfo{author}{S.~Ren},
  \bibinfo{author}{J.~Sun},
\newblock \bibinfo{title}{Deep residual learning for image recognition},
\newblock in: \bibinfo{booktitle}{Proceedings of the IEEE conference on
  computer vision and pattern recognition}, \bibinfo{year}{2016}, pp.
  \bibinfo{pages}{770--778}.
\bibitem[{Gao et~al.(2019)Gao, Yuan, Wang, and Ji}]{gao2019pixel}
\bibinfo{author}{H.~Gao}, \bibinfo{author}{H.~Yuan}, \bibinfo{author}{Z.~Wang},
  \bibinfo{author}{S.~Ji},
\newblock \bibinfo{title}{Pixel transposed convolutional networks},
\newblock \bibinfo{journal}{IEEE transactions on pattern analysis and machine
  intelligence} \bibinfo{volume}{42} (\bibinfo{year}{2019})
  \bibinfo{pages}{1218--1227}.
\bibitem[{Huang et~al.(2017)Huang, Liu, Van Der~Maaten, and
  Weinberger}]{huang2017densely}
\bibinfo{author}{G.~Huang}, \bibinfo{author}{Z.~Liu}, \bibinfo{author}{L.~Van
  Der~Maaten}, \bibinfo{author}{K.~Q. Weinberger},
\newblock \bibinfo{title}{Densely connected convolutional networks},
\newblock in: \bibinfo{booktitle}{Proceedings of the IEEE conference on
  computer vision and pattern recognition}, \bibinfo{year}{2017}, pp.
  \bibinfo{pages}{4700--4708}.
\bibitem[{Nair and Hinton(2010)}]{nair2010rectified}
\bibinfo{author}{V.~Nair}, \bibinfo{author}{G.~E. Hinton},
  \bibinfo{title}{{Rectified Linear Units Improve Restricted Boltzmann
  Machines}}, \bibinfo{year}{2010}.
\bibitem[{Maas et~al.(2013)Maas, Hannun, and Ng}]{maas2013rectifier}
\bibinfo{author}{A.~L. Maas}, \bibinfo{author}{A.~Y. Hannun},
  \bibinfo{author}{A.~Y. Ng}, \bibinfo{title}{{Rectifier NONLINEarities Improve
  Neural Network Acoustic Models}}, \bibinfo{year}{2013}.
\bibitem[{Mastromichalakis(2020)}]{mastromichalakis2020alrelu}
\bibinfo{author}{S.~Mastromichalakis},
\newblock \bibinfo{title}{Alrelu: A different approach on leaky relu activation
  function to improve neural networks performance},
\newblock \bibinfo{journal}{arXiv preprint arXiv:2012.07564}
  (\bibinfo{year}{2020}).
\bibitem[{He et~al.(2015)He, Zhang, Ren, and Sun}]{he2015delving}
\bibinfo{author}{K.~He}, \bibinfo{author}{X.~Zhang}, \bibinfo{author}{S.~Ren},
  \bibinfo{author}{J.~Sun}, \bibinfo{title}{{Delving Deep into Rectifiers:
  Surpassing Human-Level Performance on Imagenet classification}},
  \bibinfo{year}{2015}.
\bibitem[{Wang et~al.(2020)Wang, Muhammad, Hong, Sangaiah, and
  Zhang}]{wang2020alcoholism}
\bibinfo{author}{S.-H. Wang}, \bibinfo{author}{K.~Muhammad},
  \bibinfo{author}{J.~Hong}, \bibinfo{author}{A.~K. Sangaiah},
  \bibinfo{author}{Y.-D. Zhang},
\newblock \bibinfo{title}{Alcoholism identification via convolutional neural
  network based on parametric relu, dropout, and batch normalization},
\newblock \bibinfo{journal}{Neural Computing and Applications}
  \bibinfo{volume}{32} (\bibinfo{year}{2020}) \bibinfo{pages}{665--680}.
\bibitem[{Clevert et~al.(2015)Clevert, Unterthiner, and
  Hochreiter}]{clevert2015fast}
\bibinfo{author}{D.-A. Clevert}, \bibinfo{author}{T.~Unterthiner},
  \bibinfo{author}{S.~Hochreiter},
\newblock \bibinfo{title}{{Fast and Accurate Deep Network Learning by
  Exponential Linear Units (elus)}},
\newblock \bibinfo{journal}{arXiv preprint arXiv:1511.07289}
  (\bibinfo{year}{2015}).
\bibitem[{Alom et~al.(2020)Alom, Hasan, Yakopcic, Taha, and
  Asari}]{alom2020improved}
\bibinfo{author}{M.~Z. Alom}, \bibinfo{author}{M.~Hasan},
  \bibinfo{author}{C.~Yakopcic}, \bibinfo{author}{T.~M. Taha},
  \bibinfo{author}{V.~K. Asari},
\newblock \bibinfo{title}{Improved inception-residual convolutional neural
  network for object recognition},
\newblock \bibinfo{journal}{Neural Computing and Applications}
  \bibinfo{volume}{32} (\bibinfo{year}{2020}) \bibinfo{pages}{279--293}.
\bibitem[{Ruder(2016)}]{ruder2016overview}
\bibinfo{author}{S.~Ruder},
\newblock \bibinfo{title}{An overview of gradient descent optimization
  algorithms},
\newblock \bibinfo{journal}{arXiv preprint arXiv:1609.04747}
  (\bibinfo{year}{2016}).
\bibitem[{Kingma and Ba(2014)}]{kingma2014adam}
\bibinfo{author}{D.~P. Kingma}, \bibinfo{author}{J.~Ba},
\newblock \bibinfo{title}{{Adam: A Method for Stochastic Optimization}},
\newblock \bibinfo{journal}{arXiv preprint arXiv:1412.6980}
  (\bibinfo{year}{2014}).
\bibitem[{Duchi et~al.(2011)Duchi, Hazan, and Singer}]{duchi2011adaptive}
\bibinfo{author}{J.~Duchi}, \bibinfo{author}{E.~Hazan},
  \bibinfo{author}{Y.~Singer},
\newblock \bibinfo{title}{Adaptive subgradient methods for online learning and
  stochastic optimization.},
\newblock \bibinfo{journal}{Journal of machine learning research}
  \bibinfo{volume}{12} (\bibinfo{year}{2011}).
\bibitem[{Zeiler(2012)}]{zeiler2012adadelta}
\bibinfo{author}{M.~D. Zeiler},
\newblock \bibinfo{title}{Adadelta: an adaptive learning rate method},
\newblock \bibinfo{journal}{arXiv preprint arXiv:1212.5701}
  (\bibinfo{year}{2012}).
\bibitem[{Loizou and Richt{\'a}rik(2020)}]{loizou2020momentum}
\bibinfo{author}{N.~Loizou}, \bibinfo{author}{P.~Richt{\'a}rik},
\newblock \bibinfo{title}{Momentum and stochastic momentum for stochastic
  gradient, newton, proximal point and subspace descent methods},
\newblock \bibinfo{journal}{Computational Optimization and Applications}
  \bibinfo{volume}{77} (\bibinfo{year}{2020}) \bibinfo{pages}{653--710}.
\bibitem[{Dozat(2016)}]{dozat2016incorporating}
\bibinfo{author}{T.~Dozat},
\newblock \bibinfo{title}{Incorporating nesterov momentum into adam}
  (\bibinfo{year}{2016}).
\bibitem[{Cui et~al.(2018)Cui, Xue, Cai, Cao, Wang, and
  Chen}]{cui2018detection}
\bibinfo{author}{Z.~Cui}, \bibinfo{author}{F.~Xue}, \bibinfo{author}{X.~Cai},
  \bibinfo{author}{Y.~Cao}, \bibinfo{author}{G.-g. Wang},
  \bibinfo{author}{J.~Chen},
\newblock \bibinfo{title}{{Detection of Malicious Code Variants Based on Deep
  Learning}},
\newblock \bibinfo{journal}{IEEE Transactions on Industrial Informatics}
  \bibinfo{volume}{14} (\bibinfo{year}{2018}) \bibinfo{pages}{3187--3196}.
\bibitem[{Rezende et~al.(2018)Rezende, Ruppert, Carvalho, Theophilo, Ramos, and
  de~Geus}]{rezende2018malicious}
\bibinfo{author}{E.~Rezende}, \bibinfo{author}{G.~Ruppert},
  \bibinfo{author}{T.~Carvalho}, \bibinfo{author}{A.~Theophilo},
  \bibinfo{author}{F.~Ramos}, \bibinfo{author}{P.~de~Geus},
\newblock \bibinfo{title}{{Malicious Software Classification using VGG16 Deep
  Neural Network’s Bottleneck Features}},
\newblock in: \bibinfo{booktitle}{Information Technology-New Generations},
  \bibinfo{publisher}{Springer}, \bibinfo{year}{2018}, pp.
  \bibinfo{pages}{51--59}.
\bibitem[{Cui et~al.(2019)Cui, Du, Wang, Cai, and Zhang}]{cui2019malicious}
\bibinfo{author}{Z.~Cui}, \bibinfo{author}{L.~Du}, \bibinfo{author}{P.~Wang},
  \bibinfo{author}{X.~Cai}, \bibinfo{author}{W.~Zhang},
\newblock \bibinfo{title}{{Malicious Code Detection Based on CNNs and
  Multi-Objective Algorithm}},
\newblock \bibinfo{journal}{Journal of Parallel and Distributed Computing}
  \bibinfo{volume}{129} (\bibinfo{year}{2019}) \bibinfo{pages}{50--58}.
\bibitem[{Rezende et~al.(2017)Rezende, Ruppert, Carvalho, Ramos, and
  De~Geus}]{rezende2017malicious}
\bibinfo{author}{E.~Rezende}, \bibinfo{author}{G.~Ruppert},
  \bibinfo{author}{T.~Carvalho}, \bibinfo{author}{F.~Ramos},
  \bibinfo{author}{P.~De~Geus}, \bibinfo{title}{{Malicious Software
  Classification using Transfer Learning of Resnet-50 Deep Neural Network}},
  \bibinfo{year}{2017}.
\bibitem[{Vasan et~al.(2020)Vasan, Alazab, Wassan, Naeem, Safaei, and
  Zheng}]{vasan2020imcfn}
\bibinfo{author}{D.~Vasan}, \bibinfo{author}{M.~Alazab},
  \bibinfo{author}{S.~Wassan}, \bibinfo{author}{H.~Naeem},
  \bibinfo{author}{B.~Safaei}, \bibinfo{author}{Q.~Zheng},
\newblock \bibinfo{title}{{IMCFN: Image-Based Malware Classification using
  Fine-Tuned Convolutional Neural Network architecture}},
\newblock \bibinfo{journal}{Computer Networks} \bibinfo{volume}{171}
  (\bibinfo{year}{2020}) \bibinfo{pages}{107138}.
\bibitem[{Yue(2017)}]{yue2017imbalanced}
\bibinfo{author}{S.~Yue},
\newblock \bibinfo{title}{{Imbalanced Malware Images Classification: a CNN
  Based approach}},
\newblock \bibinfo{journal}{arXiv preprint arXiv:1708.08042}
  (\bibinfo{year}{2017}).
\bibitem[{Gibert et~al.(2019)Gibert, Mateu, Planes, and
  Vicens}]{gibert2019using}
\bibinfo{author}{D.~Gibert}, \bibinfo{author}{C.~Mateu},
  \bibinfo{author}{J.~Planes}, \bibinfo{author}{R.~Vicens},
\newblock \bibinfo{title}{{Using Convolutional Neural Networks for
  Classification of Malware Represented as Images}},
\newblock \bibinfo{journal}{Journal of Computer Virology and Hacking
  Techniques} \bibinfo{volume}{15} (\bibinfo{year}{2019})
  \bibinfo{pages}{15--28}.
\bibitem[{Bakour and {\"U}nver(2020)}]{bakour2020visdroid}
\bibinfo{author}{K.~Bakour}, \bibinfo{author}{H.~M. {\"U}nver},
\newblock \bibinfo{title}{{VisDroid: Android Malware Classification Based on
  Local and Global Image Features, Bag of Visual Words and Machine Learning
  Techniques}},
\newblock \bibinfo{journal}{Neural Computing and Applications}
  (\bibinfo{year}{2020}) \bibinfo{pages}{1--21}.
\bibitem[{Yang and Wen(2017)}]{yang2017detecting}
\bibinfo{author}{M.~Yang}, \bibinfo{author}{Q.~Wen}, \bibinfo{title}{{Detecting
  Android Malware by Applying Classification Techniques on Images Patterns}},
  \bibinfo{year}{2017}.
\bibitem[{Azab and Khasawneh(2020)}]{Azab9104999}
\bibinfo{author}{A.~Azab}, \bibinfo{author}{M.~Khasawneh},
\newblock \bibinfo{title}{Msic: Malware spectrogram image classification},
\newblock \bibinfo{journal}{IEEE Access} \bibinfo{volume}{8}
  (\bibinfo{year}{2020}) \bibinfo{pages}{102007--102021}.
\bibitem[{Khan et~al.(2019)Khan, Zhang, and Kumar}]{khan2019analysis}
\bibinfo{author}{R.~U. Khan}, \bibinfo{author}{X.~Zhang},
  \bibinfo{author}{R.~Kumar},
\newblock \bibinfo{title}{Analysis of resnet and googlenet models for malware
  detection},
\newblock \bibinfo{journal}{Journal of Computer Virology and Hacking
  Techniques} \bibinfo{volume}{15} (\bibinfo{year}{2019})
  \bibinfo{pages}{29--37}.
\bibitem[{Su et~al.(2018)Su, Vasconcellos, Prasad, Sgandurra, Feng, and
  Sakurai}]{Su8377943}
\bibinfo{author}{J.~Su}, \bibinfo{author}{D.~V. Vasconcellos},
  \bibinfo{author}{S.~Prasad}, \bibinfo{author}{D.~Sgandurra},
  \bibinfo{author}{Y.~Feng}, \bibinfo{author}{K.~Sakurai},
\newblock \bibinfo{title}{Lightweight classification of iot malware based on
  image recognition},
\newblock in: \bibinfo{booktitle}{2018 IEEE 42nd Annual Computer Software and
  Applications Conference (COMPSAC)}, volume~\bibinfo{volume}{02},
  \bibinfo{year}{2018}, pp. \bibinfo{pages}{664--669}.
  \DOIprefix\doi{10.1109/COMPSAC.2018.10315}.

\end{thebibliography}
\end{document}